\documentclass[11pt,twocolumn,tighten]{aastex63}
\usepackage{hyperref}
\usepackage{color}
\usepackage{url}
\usepackage{natbib}
\usepackage{graphicx}
\usepackage{xfrac}
\usepackage{ulem}
\usepackage{CJK}
\usepackage{threeparttable}
\usepackage{multirow}
\usepackage{amsmath}
\newcommand{\oii}{[\ion{O}{2}]$\lambda\lambda 3726,3729$ }
\newcommand{\deltachisq}{{\ensuremath{\Delta \chi^2}} }
\newcommand{\snroii}{\ensuremath{{\rm SNR}(F_{\rm O\,II})} }

\begin{document}

\shorttitle{[OII] profiles, star formation, and morphology of DESI ELGs}
 \shortauthors{Lan et al.}
 
\title{DESI Emission Line Galaxies:\\ Unveiling the Diversity of [OII] Profiles and its Links to Star Formation and Morphology} 

\author[0000-0001-8857-7020]{Ting-Wen Lan}
\affiliation{Graduate Institute of Astrophysics and Department of Physics, National Taiwan University, No. 1, Sec. 4, Roosevelt Rd., Taipei 10617, Taiwan}

\author{J. Xavier Prochaska}
\affiliation{Department of Astronomy and Astrophysics, University of California, Santa Cruz, 1156 High Street, Santa Cruz, CA 95065, USA}
\affiliation{Kavli Institute for the Physics and Mathematics of the Universe (Kavli IPMU), WPI, The University of Tokyo Institutes for Advanced Study (UTIAS), The
University of Tokyo, Kashiwa, Chiba, Kashiwa 277-8583, Japan}
\affiliation{Division of Science, National Astronomical Observatory of Japan, 2-21-1, Osawa, Mitaka, Tokyo 181-8588, Japan}

\author[0000-0002-2733-4559]{John Moustakas}
\affiliation{Department of Physics and Astronomy, Siena College, 515 Loudon Road, Loudonville, NY 12211, USA}

\author[0000-0002-2949-2155]{Małgorzata Siudek}
\affiliation{Institute of Space Sciences, ICE-CSIC, Campus UAB, Carrer de Can Magrans s/n, 08913 Bellaterra, Barcelona, Spain}
\affiliation{Instituto Astrofisica de Canarias, Av. Via Lactea s/n, E38205 La Laguna, Spain
}

\author{J.~Aguilar}
\affiliation{Lawrence Berkeley National Laboratory, 1 Cyclotron Road, Berkeley, CA 94720, USA}
\author[0000-0001-6098-7247]{S.~Ahlen}
\affiliation{Physics Dept., Boston University, 590 Commonwealth Avenue, Boston, MA 02215, USA}
\author[0000-0001-9712-0006]{D.~Bianchi}
\affiliation{Dipartimento di Fisica ``Aldo Pontremoli'', Universit\`a degli Studi di Milano, Via Celoria 16, I-20133 Milano, Italy}
\author{D.~Brooks}
\affiliation{Department of Physics \& Astronomy, University College London, Gower Street, London, WC1E 6BT, UK}
\author{T.~Claybaugh}
\affiliation{Lawrence Berkeley National Laboratory, 1 Cyclotron Road, Berkeley, CA 94720, USA}
\author[0000-0002-5954-7903]{S.~Cole}
\affiliation{Institute for Computational Cosmology, Department of Physics, Durham University, South Road, Durham DH1 3LE, UK}

\author{K.~Dawson}
\affiliation{Department of Physics and Astronomy, The University of Utah, 115 South 1400 East, Salt Lake City, UT 84112, USA}
\author[0000-0002-1769-1640]{A.~de la Macorra}
\affiliation{Instituto de F\'{\i}sica, Universidad Nacional Aut\'{o}noma de M\'{e}xico,  Cd. de M\'{e}xico  C.P. 04510,  M\'{e}xico}
\author{P.~Doel}
\affiliation{Department of Physics \& Astronomy, University College London, Gower Street, London, WC1E 6BT, UK}
\author[0000-0002-2890-3725]{J.~E.~Forero-Romero}
\affiliation{Departamento de F\'isica, Universidad de los Andes, Cra. 1 No. 18A-10, Edificio Ip, CP 111711, Bogot\'a, Colombia}
\affiliation{Observatorio Astron\'omico, Universidad de los Andes, Cra. 1 No. 18A-10, Edificio H, CP 111711 Bogot\'a, Colombia}
\author{E.~Gaztañaga}
\affiliation{Institut d'Estudis Espacials de Catalunya (IEEC), 08034 Barcelona, Spain}
\affiliation{Institute of Cosmology and Gravitation, University of Portsmouth, Dennis Sciama Building, Portsmouth, PO1 3FX, UK}
\affiliation{Institute of Space Sciences, ICE-CSIC, Campus UAB, Carrer de Can Magrans s/n, 08913 Bellaterra, Barcelona, Spain}

\author[0000-0003-3142-233X]{S.~Gontcho A Gontcho}
\affiliation{Lawrence Berkeley National Laboratory, 1 Cyclotron Road, Berkeley, CA 94720, USA}

\author{G.~Gutierrez}
\affiliation{Fermi National Accelerator Laboratory, PO Box 500, Batavia, IL 60510, USA}
\author[0000-0001-9822-6793]{J.~Guy}
\affiliation{Lawrence Berkeley National Laboratory, 1 Cyclotron Road, Berkeley, CA 94720, USA}
\author{K.~Honscheid}
\affiliation{Center for Cosmology and AstroParticle Physics, The Ohio State University, 191 West Woodruff Avenue, Columbus, OH 43210, USA}
\affiliation{Department of Physics, The Ohio State University, 191 West Woodruff Avenue, Columbus, OH 43210, USA}
\affiliation{The Ohio State University, Columbus, 43210 OH, USA}
\author{R.~Kehoe}
\affiliation{Department of Physics, Southern Methodist University, 3215 Daniel Avenue, Dallas, TX 75275, USA}
\author[0000-0003-3510-7134]{T.~Kisner}
\affiliation{Lawrence Berkeley National Laboratory, 1 Cyclotron Road, Berkeley, CA 94720, USA}
\author{A.~Lambert}
\affiliation{Lawrence Berkeley National Laboratory, 1 Cyclotron Road, Berkeley, CA 94720, USA}
\author[0000-0003-1838-8528]{M.~Landriau}
\affiliation{Lawrence Berkeley National Laboratory, 1 Cyclotron Road, Berkeley, CA 94720, USA}
\author[0000-0002-1125-7384]{A.~Meisner}
\affiliation{NSF NOIRLab, 950 N. Cherry Ave., Tucson, AZ 85719, USA}
\author{R.~Miquel}
\affiliation{Instituci\'{o} Catalana de Recerca i Estudis Avan\c{c}ats, Passeig de Llu\'{\i}s Companys, 23, 08010 Barcelona, Spain}
\affiliation{Institut de F\'{i}sica d’Altes Energies (IFAE), The Barcelona Institute of Science and Technology, Campus UAB, 08193 Bellaterra Barcelona, Spain}

\author{A.~Muñoz-Gutiérrez}
\affiliation{Instituto de F\'{\i}sica, Universidad Nacional Aut\'{o}noma de M\'{e}xico,  Cd. de M\'{e}xico  C.P. 04510,  M\'{e}xico}

\author[0000-0001-8684-2222]{J.~ A.~Newman}
\affiliation{Department of Physics \& Astronomy and Pittsburgh Particle Physics, Astrophysics, and Cosmology Center (PITT PACC), University of Pittsburgh, 3941 O'Hara Street, Pittsburgh, PA 15260, USA}
\author{C.~Poppett}
\affiliation{Lawrence Berkeley National Laboratory, 1 Cyclotron Road, Berkeley, CA 94720, USA}
\affiliation{Space Sciences Laboratory, University of California, Berkeley, 7 Gauss Way, Berkeley, CA  94720, USA}
\affiliation{University of California, Berkeley, 110 Sproul Hall \#5800 Berkeley, CA 94720, USA}
\author[0000-0001-7145-8674]{F.~Prada}
\affiliation{Instituto de Astrof\'{i}sica de Andaluc\'{i}a (CSIC), Glorieta de la Astronom\'{i}a, s/n, E-18008 Granada, Spain}
\author{G.~Rossi}
\affiliation{Department of Physics and Astronomy, Sejong University, Seoul, 143-747, Korea}
\author[0000-0002-9646-8198]{E.~Sanchez}
\affiliation{CIEMAT, Avenida Complutense 40, E-28040 Madrid, Spain}

\author{M.~Schubnell}
\affiliation{Department of Physics, University of Michigan, Ann Arbor, MI 48109, USA}
\affiliation{University of Michigan, Ann Arbor, MI 48109, USA}

\author[0000-0002-6588-3508]{H.~Seo}
\affiliation{Department of Physics \& Astronomy, Ohio University, Athens, OH 45701, USA}

\author{D.~Sprayberry}
\affiliation{NSF NOIRLab, 950 N. Cherry Ave., Tucson, AZ 85719, USA}

\author[0000-0003-1704-0781]{G.~Tarl\'{e}}
\affiliation{University of Michigan, Ann Arbor, MI 48109, USA}

\author{B.~A.~Weaver}
\affiliation{NSF NOIRLab, 950 N. Cherry Ave., Tucson, AZ 85719, USA}

\author[0000-0002-6684-3997]{H.~Zou}
\affiliation{National Astronomical Observatories, Chinese Academy of Sciences, A20 Datun Rd., Chaoyang District, Beijing, 100012, P.R. China}

\begin{abstract}
We study the [OII] profiles of emission line galaxies (ELGs) from the Early Data Release of the Dark Energy Spectroscopic Instrument (DESI). To this end, we decompose and classify the shape of [OII] profiles with the first two eigenspectra derived from Principal Component Analysis. 
Our results show that DESI ELGs have diverse line profiles which can be categorized into three main types: (1) narrow lines with a median width of $\sim50$ km/s, (2) broad lines with a median width of $\sim80$ km/s, and (3) two-redshift systems with a median velocity separation of $\sim 150$ km/s, i.e., double-peak galaxies. To investigate the connections between the line profiles and galaxy properties, we utilize the information from the COSMOS dataset and compare the properties of ELGs, including star-formation rate (SFR) and galaxy morphology, with the average properties of reference star-forming galaxies with similar stellar mass, sizes, and redshifts. Our findings show that on average, DESI ELGs have higher SFR and more asymmetrical/disturbed morphology than the reference galaxies. 
Moreover, we uncover a relationship between the line profiles, the excess SFR and the excess asymmetry parameter, showing that DESI ELGs with broader [OII] line profiles have more disturbed morphology and higher SFR than the reference star-forming galaxies. Finally, we discuss possible physical mechanisms giving rise to the observed relationship and the implications of our findings on the galaxy clustering measurements, including the halo occupation distribution modeling of DESI ELGs and the observed excess velocity dispersion of the satellite ELGs.

\end{abstract}
\section{Introduction}
Galaxies that produce strong emission lines from the HII star-forming regions, i.e. emission line galaxies (ELGs), have become one of the key observational targets for large cosmological surveys. Via the strong emission line features, one can detect and measure the redshifts of ELGs across a wide range of cosmic time with accessible observational resources and use those galaxies as tracers of the 3D large-scale structure of the Universe. Previous cosmological surveys, including the WiggleZ Dark Energy Survey \citep{Drinkwater2010} and the Extended Baryon Oscillation Spectroscopic Survey in the Sloan Digital Sky Survey-IV \citep[SDSS-IV eBOSS,][]{DawsoneBOSS} projects, have demonstrated that cosmological parameters, such as the expansion rate of the Universe and the dark energy equation of state, can be constrained via the baryonic acoustic oscillations (BAOs) \citep[e.g.,][]{Eisenstein2005} signals detected from the ELG clustering measurements \citep[e.g.,][]{Blake2011, Raichoor2021}. The success of these programs has motivated the ongoing and upcoming surveys, including the Dark Energy Spectroscopic Instrument \citep[DESI,][]{Levi2013, DESI2016, DESI2016b, DESI_overview}, the Prime-Focus Spectrograph \citep[PFS,][]{Takada2014}, and \textit{Euclid} \citep{Euclid2022} to select ELGs as one of the main targets and collect tens of millions of spectroscopic measurements of ELGs for their cosmological programs. 

Among the emission lines produced by HII regions, the \oii doublet has been considered as one of the key transitions for ground-based spectroscopic measurements \citep[e.g.,][]{Comparat2015} owing to its detectibility in optical and near-infrared wavelengths from the local Universe to redshift $\sim2$, the cosmic epoch at the peak of the star-formation rate density \citep[see][for a review]{Madau2014}. 
The \oii doublet can be used to not only pinpoint the redshifts of galaxies but also obtain information of galaxy physical properties, including star-formation rate (line luminosity), electron density (doublet line ratio), and gas kinematics (line width) \citep[e.g.,][]{Kennicutt1998, Moustakas2006, OsterbrockAGN,Kassin2012,Kewley2019}. 
However, to extract such information from spectra requires a spectral resolution, $R\sim3300$, to resolve the \oii doublet feature \citep{Comparat2013}. This limits the exploration and characterization of the [OII] line properties of ELGs with the spectra from previous cosmological surveys, such as WiggleZ \citep[$R\sim1300$,][]{Drinkwater2010} and SDSS-IV eBOSS \citep[$R\sim2000$,][]{DawsoneBOSS}. On the other hand, the DESI survey has the required spectral resolution \citep[][]{DESI_overview} and will compile a large ELG catalog with resolved [OII] lines from redshift 0.6 to redshift 1.6, a redshift region rarely probed to date. With this large spectroscopic dataset, it is crucial to understand the underlying galaxy population selected as ELGs in the DESI survey and characterize their physical properties to fully utilize this big dataset and maximize the scientific returns for cosmology and galaxy science.

In this work, we explore diversity in the DESI ELG galaxy population based on their [OII] emission line profiles and investigate the relationships between the line profiles and the physical properties of the galaxies. 
To this end, we make use of a principal component analysis \citep[PCA,][for a review]{PCA_review}, a dimensional reduction technique, to obtain the key eigen-spectra, which carry physical information of the line profiles. We use them to describe line profiles and classify galaxies.
We then obtain the physical properties, such as stellar mass ($\rm M_{*}$), star-formation rate (SFR), and morphology parameters, of a subset of ELGs in the COSMOS field \citep{Scoville2007b} and examine the connections between line profiles and galaxy properties. This enables us to reveal possible underlying mechanisms driving the [OII] line profiles. 

The structure of the paper is as follows. The datasets used in this analysis are described in Section 2. We present the PCA results in Section 3 and explore the relations between line profiles and galaxy properties in Section 4. We discuss our results and their implications in Section 5, and summarize in Section 6.
Throughout the paper, we adopt a flat $\rm \Lambda$CDM cosmology with
$h = 0.694$ and $\rm \Omega_{m}=0.287$ \citep[WMAP9][]{WMAP9}. We use AB magnitudes corrected with Galactic extinction \citep{Schlegel1998}.
\section{Datasets}
\subsection{DESI ELG spectra}
The Dark Energy Spectroscopic Instrument (DESI) survey \citep{DESI_overview} is a Stage-IV project, primarily designed for probing the cosmological parameters of the Universe \citep{Levi2013, DESI2016, DESI_SV}. DESI consists of bright-time and dark-time science targets \citep{DESI_MW, DESI_BGS,DESI_LRG,DESI_ELG,DESI_QSO} with 
[OII] emission line galaxies \citep[ELGs,][]{DESI_ELG} being one of the key targets in the program. In order to obtain the desired redshift range of DESI ELGs, a specific target selection scheme is developed for the sources detected in the DESI Legacy Imaging Surveys \citep[][]{Dey2019},  which include $g$, $r$, $z$ bands and near-infrared images from WISE satellite \citep{Wright2010}, via the Tractor algorithm \citep{Lang2016} with large galaxies masked \citep{Moustakas2023}. To ensure the efficiency of DESI operations, dedicated pipelines are developed for target selection \citep{Myers2023}, survey operations \citep{Schlafly2023} and fiber assignment \citep{Raichoor2024}.

Spectroscopic data is collected by the DESI instrumentation installed on the 4-meter Mayall Telescope at Kitt Peak National Observatory \citep{DESI_overview,Silber2023, Miller2023}. The instrument has a 3.2 degree diameter field of view covered with 5020 robotic fiber positioners. The wavelength coverage ranges from 3600 $\rm \AA$ to 9800 $\rm \AA$ with spectral resolution from $\sim2000$ in the shortest wavelength to $\sim5000$ in the longest wavelength. The raw spectra are processed, reduced and calibrated by an automatic pipeline \citep{Guy2023}. The redshifts of sources are then obtained via an algorithm, called $Redrock$ \citep{Bailey2023} \citep[See also][]{Brodzeller2023, Anand2024}, with the redshift accuracy and precision for all types of extragalactic sources, including ELGs, being tested and validated against redshifts obtained from spectra with long exposure times and confirmed via visual inspection \citep{Lan2023, Alexander2023}.

In this analysis, we make use of the DESI spectra of ELGs observed during the Survey Validation phase \citep{DESI_SV} to explore the \oii emission line profiles. The DESI SV campaign consists of two key phases, the SV1 and the one-percent survey. The SV1 observations include targets selected from a wider parameter space in order to test and finalize the target selection scheme of the DESI main survey. The one-percent survey then adopted the finalized target selection scheme and obtained the observations that cover sky area $\rm \sim 140\, deg^{2}$, about $1\%$ of the DESI final footprint \citep{DESI_SV}. 

We focus on the ELGs observed in the one-percent survey to inform the properties of ELGs of the DESI main survey. 
The ELGs are selected within a color-magnitude space defined with $g-r$, $r-z$ colors, g-band magnitude and g-band fiber magnitude for preferentially including star-forming galaxies at $1.1<z<1.63$ \citep{DESI_ELG}. 

We first make use of the data from the Early Data Release \citep{DESI_EDR} which includes $\sim 280,000$ spectra of ELGs with robust redshift measurements that pass the redshift quality selection for ELGs, $\rm log_{10}\, \snroii>0.9-0.2\times log_{10} \, \deltachisq$, where $\rm \snroii$  is the signal to noise ratio of the [OII] emission lines measured via a double-gaussian fitting \citep{DESI_ELG} and \deltachisq is the difference between the $\chi^{2}$ values from the second best-fit model and the first best-fit model provided by $Redrock$ \citep{Bailey2023}. This selection yields a ELG sample with redshift purity (the fraction of ELGs with $\it{Redrock}$ redshifts and visually inspected redshifts difference smaller than $\sim1000$ km/s) greater than $99\%$ \citep{Lan2023}. We further select ELGs with $0.6<z<1.63$ and with spectral types identified by $\it{Redrock}$ as galaxies. This selection yields a sample of $\sim 260,000$ ELGs.  In addition, using line information available at different redshifts, we remove possible active galactic nuclei (AGN) contamination. The selections remove approximately $4\%$ of DESI ELGs. The detailed selections are described in Appendix~\ref{appendix:AGN}. 
Approximately $250,000$ ELGs pass the above selection criteria. 
We note that this sample includes ELGs with LOP (Low Priority) and VLO (Very Low Priority) selections \citep[See][for details]{DESI_ELG} for the main DESI survey. Approximately 78\% of the sample is LOP ELGs, the fiducial ELG sample for DESI clustering measurements.

\subsection{The COSMOS datasets}
To explore the physical properties of DESI ELGs in the context of the whole star-forming galaxy population across redshifts, we utilize the COSMOS2020 catalog \citep{COSMOS2020} which includes approximately 1.7 million sources with photometric redshifts ($z_{photo}$) and physical properties of galaxies estimated based on multi-wavelength  broad-band deep imaging datasets. The COSMOS field was covered by the DESI one-percent survey. Therefore, we can compile a sample with a few thousand DESI ELGs with properties derived from deep COSMOS datasets.

\textbf{Catalog selection:} In the COSMOS2020 data release, there are two source catalogs, the {\it CLASSIC} catalog and the {\it FARMER} catalog, which are constructed based on the {\it SExtractor} \citep{Bertin1996} and the {\it Tractor} \citep{Lang2016} algorithms respectively for source detection and flux measurements. The primary difference between two methods is that in the {\it CLASSIC} catalog, the fluxes of sources are measured with the aperture photometry method while in the {\it FARMER} catalog, the fluxes of sources are measured with the forced photometric best-fit models of source light distribution. As shown in \citet{COSMOS2020}, for sources brighter than 24 mag in 
Hyper Suprime-Cam z-band, these two methods yield consistent measurements with magnitude deviation lower than 0.05 magnitude. However, the {\it FARMER} catalog only includes sources within the UltraVISTA footprint \citep{McCracken2012}, which consists of about half the amount of sources in the {\it CLASSIC} catalog with the full COSMOS field coverage. To maximize the number of the matched DESI ELGs within the COSMOS catalog, we use the measurements provided by the {\it CLASSIC} catalog. 

\textbf{Photometric redshifts and physical properties:} The COSMOS2020 catalog also includes photometric redshifts and the estimated physical properties of galaxies, including the stellar mass ($M_{*}$) and star-formation rate (SFR), based on the LePhare \citep{Ilbert2006} and EAZY \citep{Brammer2008} codes. These two algorithms adopt different sets of galaxy spectral energy distribution (SED) templates and different star-formation histories. For the details, we refer the readers to \citet{COSMOS2020} and \citet{cosmos_eazy}.  

In this work, we adopt the photometric redshifts and galaxy properties from the EAZY algorithm. This choice is motivated by the fact that for the DESI ELGs in the COSMOS2020 footprint and with $\rm |z_{DESI}-z_{photo}|<0.1$, the SFR based on the $H\alpha$ emission line luminosity
\citep[][]{Kennicutt1998, Kennicutt2009} from the EAZY algorithm has a higher Spearman's correlation coefficient ($\rho_{s}\simeq 0.72$) with [OII] emission line luminosity, directly measured from the DESI spectra, than the estimated SFR provided by EAZY ($\rho_{s}\simeq 0.68$) and LePhare ($\rho_{s}\simeq 0.66$) codes. More specifically, we use the best-fit $z_{photo}$, the stellar mass, and the SFR based $H\alpha$ emission line luminosity from the EAZY algorithm applied to the {\it CLASSIC} catalog. We note that the COSMOS2020 catalog provides dust-reddened $H\alpha$ luminosity. We correct the dust effect by using $A_{H\alpha}=0.81 \, A_{V}$ \citep{Kriek2013} and adopt the $H\alpha$ intrinsic luminosity and SFR relation from \citet{Kennicutt2009} for the Chabrier initial mass function \citep{Chabrier2003}.
All the stellar mass and the SFR of galaxies used in the work are based on the above method.
We emphasize that the goal of this research is to explore the relationship between [OII] line profiles and physical properties of galaxies and to compare the physical properties of DESI ELGs with the properties of the overall star-forming galaxy population at similar redshifts. Identifying the most accurate estimates of physical properties of DESI ELGs is beyond the scope of this paper.   

\textbf{Morphological parameters:} Besides the photometric redshifts and physical properties, we make use of the non-parametric diagnostics of galaxy morphology provided by the Zurich Estimator of Structural Type catalog \citep[ZEST,][]{COSMOS_zbest, cosmos_ACS}. The catalog includes 
\begin{itemize}
    \item asymmetry (A), which quantifies the rotational symmetry of galaxy light distribution;
    \item concentration (C), which quantifies the concentration of galaxy light distribution;
    \item Gini coefficient (G), which quantifies the uniformity of the light distribution,
    \item M20, which is second-order moment of the brightest $20\%$ pixels and
    \item R80, which is the length of the best-fit ellipse including $80\%$ of total light along semi-major axis in unit of arcsec ("), reflecting the observed size of the galaxy, 
\end{itemize}
  \citep[e.g.,][for a review]{Abraham2003, Lotz2004, Conselice2014}. These parameters are derived from Hubble Space Telescope (\textit{HST}) Advanced Camera for Surveys (ACS) F814W images \citep[][]{Scoville2007, Koekemoer2007, Leauthaud2007} for galaxies with $I_{AB}<24$, a depth sufficient to include all the DESI ELGs \citep{DESI_ELG}.
  The F814W filter covers the rest-frame near ultraviolet to optical wavelengths of DESI ELGs. At $z\sim0.6$, the filter covers  from $\rm 4400 \, \AA$ to $\rm 5900 \, \AA$ and at $z\sim1.6$,  it covers from $\rm 2700 \, \AA$ to $\rm 3600 \, \AA$. The average width of the point spread function for the \textit{HST} ACS images is $\sim0.1"$ \citep{Koekemoer2007} which corresponds to $\sim0.67$ kpc at $z=0.6$ and $\sim0.86$ kpc at $z=1.6$.

Combining the information from these three catalogs\footnote{
Except $\sim 0.4\%$ of DESI ELGs within the locations of the stellar masks of the COSMOS2020 catalog, the rest of DESI ELGs in the COSMOS field have matched counterparts within $0.75"$ radius.}, we produce two samples of DESI ELGs with $\rm |z_{DESI}-z_{photo}|<0.1$:
\begin{itemize}
    \item \textbf{DESI-COSMOS}: The first one, DESI-COSMOS, is the combination between the DESI ELGs and COSMOS2020 {\it CLASSIC} catalog, which is done by cross-matching sources with 0.75" radius. This yields a sample with $\sim 4,500$ sources having photo-z, stellar mass, and SFR measurements. 
    \item \textbf{DESI-ZEST}: The second sample, DESI-ZEST, includes the morphological information from the ZEST catalog by matching the DESI-COSMOS sources with the ZEST sources with 0.75" radius.
    This DESI-ZEST sample is a subset of the DESI-COSMOS sample, consisting of about 2,200 galaxies.
\end{itemize}
We note that there are rare cases ($\sim0.3\%$) that one ELG has two or more galaxies within 0.75" radius from the COSMOS2020 or the ZEST catalogs. For such cases, we adopt the properties of the most massive and brightest (in F814W filter) galaxy.

In order to compare the properties of DESI ELGs and that of the overall star-forming galaxies at the similar redshifts, we construct reference samples by selecting star-forming galaxies from the COSMOS2020 catalog that satisfy
\begin{equation}
\label{eq:SFR}
   \rm log_{10}\,SFR>
   -0.65+0.65\times(log_{10}\,M_{*}-10)+1.07\times(z-0.1).
\end{equation}
This functional form is adopted from \citet[][]{Moustakas2013} for separating star-forming and quiescent galaxies with a modification on the zero point to better include star-forming galaxies based on the adopted SFR. We refer this star-forming galaxy sample from the full COSMOS2020 catalog as COSMOS-SFG. We also cross-match this COSMOS-SFG sample with the ZEST catalog with $0.75"$ radius to obtain a sample of approximately 100,000 star-forming galaxies with morphological parameters based on HST-ACS images. This sample is referred as COSMOS-ZEST. These samples will be used in Section 4. 

\section{PCA decomposition and classification}
The basic of the principal component analysis  \citep[PCA][for a review]{PCA_review} is to calculate the eigen-values of the correlation matrix for a
given dataset and the corresponding 
eigen-vectors. By doing so, one can identify key eigen-vectors that carry the bulk of the variance of the dataset. Via PCA, each original data vector can be described by the sum of the mean vector and a linear combination of the eigen-vectors. 

In this analysis, our data matrix consists of DESI ELG spectra around [OII] emission line region ($\rm 3722<\lambda<3735 \, \AA$). Via the PCA decomposition, each ELG spectrum can be described as 
\begin{equation}
    f_{\lambda, i} = \langle \mu \rangle+ \sum_{j=1}^{m} \rm Coeff_{\it ij} \times |\xi_{\it j}\rangle, 
\end{equation}
where $f_{\lambda, i}$ is the $i$th ELG spectrum, $\langle \mu \rangle$ is the mean spectrum of the whole ELG spectra, $|\xi_{\it j}\rangle$ is $j$th eigen-vector (eigen-spectrum), and $\rm Coeff_{ij}$ is the coefficient for $i$th ELG spectrum and $j$th eigen-spectrum. In this manner, one 
describes each spectrum with only a few coefficients and may explore and classify [OII] profiles based on their values. 
This approach has been adopted in previous studies for exploring various types of astronomical datasets, such as quasar spectra, galaxy spectra, and galaxy physical properties \citep[e.g.,][]{SuzukiPCA, YipPCA, PCAgalaxu}. In this work, we use the weighted PCA algorithm, which takes into account the measurement uncertainty and missing data, developed by \citet{Delchambre2015}\footnote{We use the package at \url{https://github.com/jakevdp/wpca} developed by Jake VanderPlas.}.

\subsection{Spectral Processing}
In order to perform PCA, we shift all the ELG spectra to their rest-frame based on their $\it Redrock$ redshifts and project the spectra on to a common wavelength grid with a pixel size $\rm 0.5\,\AA$ ($\sim40$ km/s). 
This pixel size corresponds to the original pixel size of DESI spectra ($\rm 0.8\,\AA$) in the rest-frame of a source at redshift 0.6, the lower bound of the redshift in our analysis. 
We estimate the continuum of each spectrum using the median value of pixels around \oii emission lines ($\rm 3715<\lambda<3722 \, \AA$ and $\rm 3735 <\lambda<3742 \, \AA$) and subtract the estimated continuum from the spectrum. 
We then normalize each spectrum by the maximum value of the pixels with $\rm S/N>3$ between $3726\, \rm \AA$ and $3731\, \rm \AA$ of the spectrum. By doing so, the results of PCA is more sensitive to the shape variation of the \oii emission lines, the focus of this study, than the flux difference between spectra. Finally, we remove spectra with any bad pixels and with the median S/N of the [OII] regions between $3726\, \rm \AA$ and $3731\, \rm \AA$ lower than 2. The final sample includes 229,996 ELGs. 

\begin{figure}
\center
\includegraphics[width=0.44\textwidth]{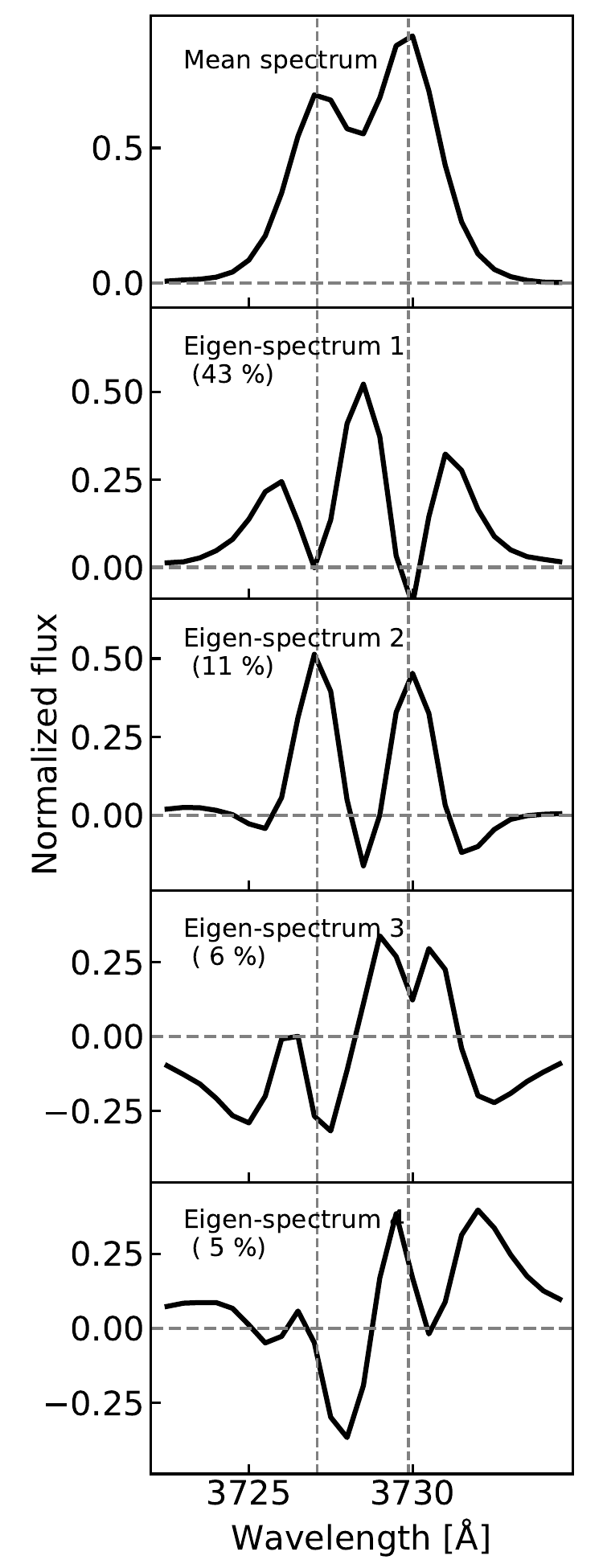}
\caption{Mean spectrum and PCA eigen-spectra of DESI ELGs. The top panel shows the mean spectrum and the 2nd to 5th panels show the eigen-spectra ordered by their explained variance listed in each panel. The two vertical dashed lines indicate the wavelengths of the [OII] doublet.}
\label{fig:eigen-spectra}
\end{figure}

\begin{figure*}
\center
\includegraphics[width=0.97\textwidth]{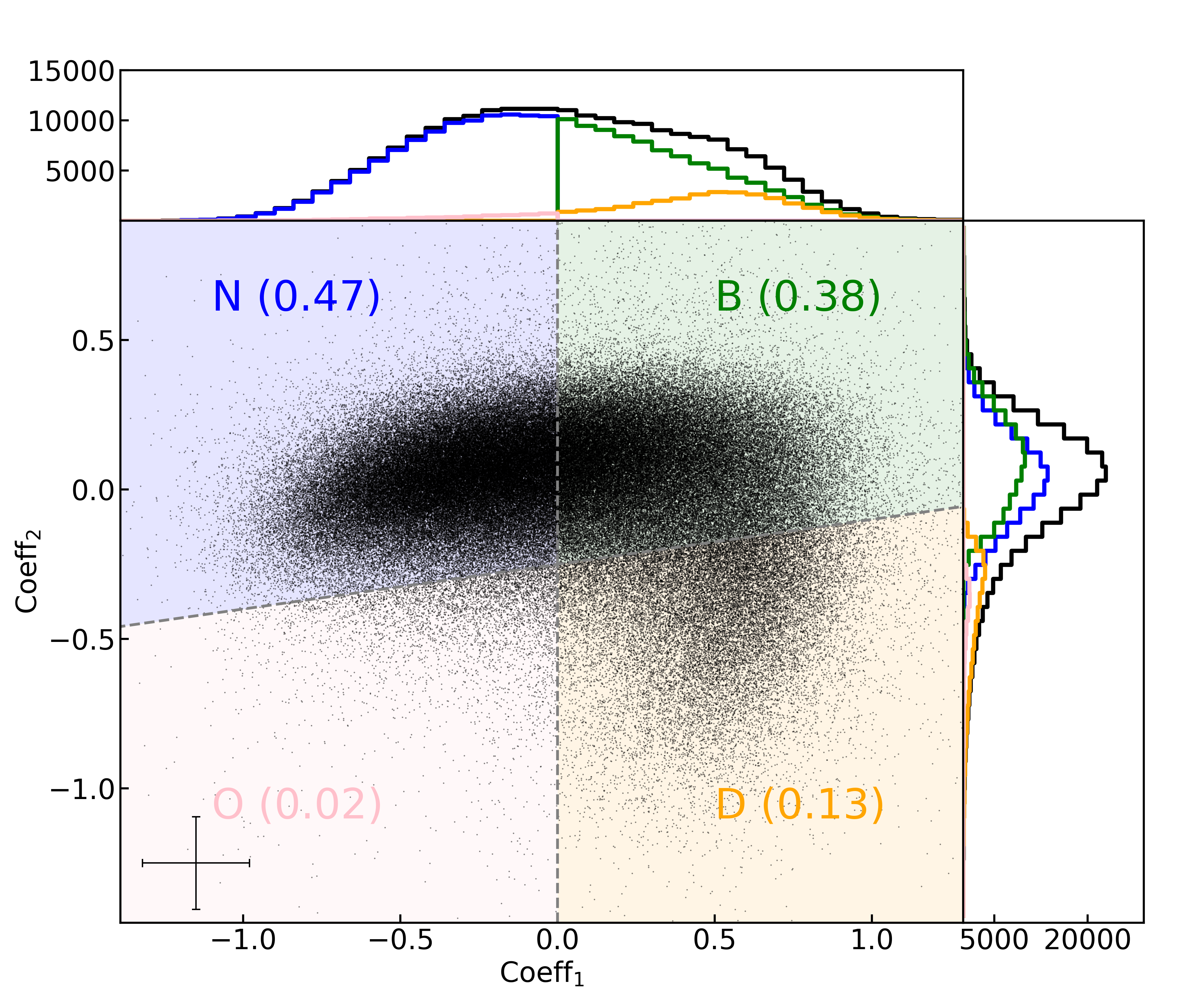}
\caption{Distribution of ELGs in the PCA coefficient 1 ($\rm Coeff_{1}$, x-axis) and 2 ($\rm Coeff_{2}$, y-axis) space. ELGs are separated into four regions, narrow (N), broad (B), double-peak (D), and outlier (O), by the two dashed lines ($\rm Coeff_{2}=0.15\times Coeff_{1}-0.25$ and $\rm Coeff_{1}=0$) and with background colors being blue, green, orange, and pink respectively. The value shown in each region indicates the fraction of total sources in the region and the top and right panels show the number distributions of the four types as a function of $\rm Coeff_{1}$ and $\rm Coeff_{2}$ respectively. 
The median values of the uncertainties of $\rm Coeff_{1}$ and $\rm Coeff_{2}$ are 0.17 and 0.16 respectively as shown in the lower left corner. The uncertainties of the PCA coefficients of each spectrum are estimated by perturbing the spectrum with the observed spectral noise 1000 times and obtaining the corresponding coefficients and the standard deviations. 
} 
\label{fig:coefficients}
\end{figure*}
\begin{figure*}
\center
\includegraphics[width=0.95\textwidth]{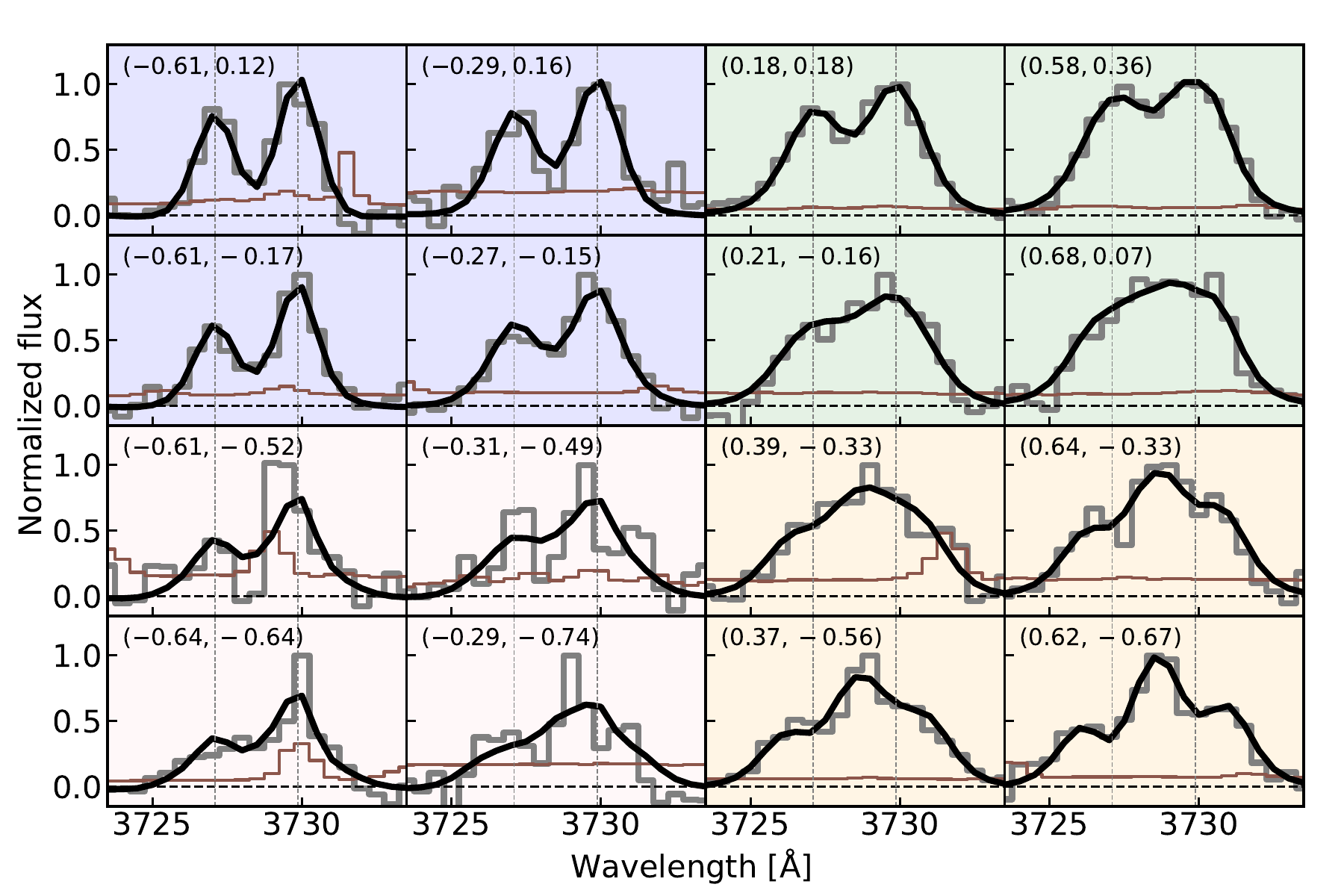}
\caption{Example spectra of DESI ELGs as a function of $\rm Coeff_{1}$ and $\rm Coeff_{2}$. The values shown on the top left corner in each panel are ($\rm Coeff_{1}$, $\rm Coeff_{2}$) values of the ELG. The grey spectrum is the observed spectrum, the black curve is the PCA reconstructed spectrum based on the first two eigen-spectra, and the uncertainty array is in brown. The background colors of the panels indicate the locations of the ELGs in the $\rm Coeff_{1}$ and $\rm Coeff_{2}$ space as shown in Figure~\ref{fig:coefficients}.} 
\label{fig:example_spectra}
\end{figure*}

\subsection{Eigen-spectra}
The mean spectrum and the eigen-spectra obtained by PCA are shown in Figure~\ref{fig:eigen-spectra}. The top panel shows the mean spectrum of the ELGs with clear \oii emission lines. The lower four panels show the first four eigen-spectra respectively. 
\begin{itemize}
    \item \textbf{1st eigen-spectrum:} the 1st eigen-spectrum explains $43\%$ of the variance of the [OII] line region. It
    has three peaks at 3726 $\rm \AA$, 3728.5 $\rm \AA$, and 3731 $\rm \AA$, which do not coincide with the \oii lines. 
    The \oii lines in the 1st eigen-spectrum are nearly zero. 
    By modulating flux off the line center, this eigen-spectrum broadens the 
    line width of the emission with positive coefficients and reduces the line width with negative coefficients.
    
    \item \textbf{2nd eigen-spectrum:} the 2nd eigen-spectrum explains $11\%$ of the variance. It has two peaks that coincide with the wavelengths 
    of \oii. With a positive value coefficient, this component increases the \oii line strength and with a negative coefficient, it decreases the strength. Therefore, the 2nd eigen-spectrum modulates the amplitude and peakedness of the \oii emission lines. 
    
    \item \textbf{3rd and 4th eigen-spectra:} the 3rd and 4th eigen-spectra both explain only $\sim 5\%$ of the variance. They both have positive values around [OII] 3730 line and negative values around [OII] 3727 line. Therefore, these two components can account for the asymmetry of the line profiles induced by both physical and non-physical signals, such as the variation of the line ratio between [OII]$\lambda3726$ and [OII]$\lambda3729$ \citep[e.g.,][]{COSMOS_OII} and the residuals of sky lines.

\end{itemize}
The first two eigen-spectra explain $\sim54\%$ (43\%, 11\%) of the total variances of the ELG spectra. In the following, we focus on the first two eigen-spectra and their application in characterizing the DESI ELG \oii line profiles, while we plan to investigate the application of the 3rd and 4th eigen-spectra in future studies.

\subsection{Coefficients}
Using only the first two eigen-spectra, the \oii line profile of each ELG can be described by two coefficients. Figure~\ref{fig:coefficients} shows the distribution of Coefficient 1 ($\rm Coeff_{1}$) (x-axis) and Coefficient 2 ($\rm Coeff_{2}$) (y-axis). We find that the \oii line profiles of DESI ELGs cover the parameter space with both coefficient 1 and coefficient 2, ranging from positive to negative values. Based on the ($\rm Coeff_{1}, \rm Coeff_{2}$) values, we classify ELGs into four regions using two empirical selection boundaries $\rm Coeff_{2}=0.15\times Coeff_{1}-0.25$ and $\rm Coeff_{1}=0$ and show example spectra in Figure~\ref{fig:example_spectra}:

\begin{itemize}
    \item \textbf{Narrow region (N):} $\rm Coeff_{1}<0$ and $\rm Coeff_{2}>0.15\times Coeff_{1} - 0.25$. In this region, the doublet of \oii is narrow and is clearly separated. 
    
    \item \textbf{Broad region (B):} $\rm Coeff_{1}>0$ and $\rm Coeff_{2}>0.15\times Coeff_{1} - 0.25$. With the increase of $\rm Coeff_{1}$ values, the \oii line width becomes broader and the two [OII] lines gradually blend together with increasing $\rm Coeff_{1}$. 
    \item \textbf{Double-peak\footnote{
    We note that the [OII] profiles have three peaks which are in fact consisted of two components with overlapping [OII] emission line doublet. For transition with a single line, the emission line profiles are double-peak. Following the literature, we use "double-peak" galaxies for describing these systems.} region (D):} $\rm Coeff_{1}>0$ and $\rm Coeff_{2}<0.15\times Coeff_{1}-0.25$. The combination of a positive value of $\rm Coeff_{1}$ and a negative value of $\rm Coeff_{2}$  yields [OII] line profiles with three emission line peaks with the central peak at $\rm \sim 3728.5\, \AA$ as shown in lower right panels of Figure~\ref{fig:example_spectra}. 
    These profiles are consistent with [OII] lines from two components with velocity offsets. 
    The two velocity component spectral features are also observed in other emission lines, such as $\rm H\beta$ and $\rm [OIII]$, for low-z ELGs with double-peak emission lines. Within the selection, there are $\simeq 30,000$ double-peak galaxies identified from the DESI EDR dataset, which is currently the largest double-peak galaxy catalog at $z>0.6$. 
    \item \textbf{Outlier region (O):} $\rm Coeff_{1}<0$ and $\rm Coeff_{2}<0.15\times Coeff_{1}-0.25$. In this region, the [OII] line profiles are affected by residuals of strong sky emission lines as shown in the lower left panels of Figure~\ref{fig:example_spectra}.
\end{itemize}

\begin{figure*}
\center
\includegraphics[width=1\textwidth]{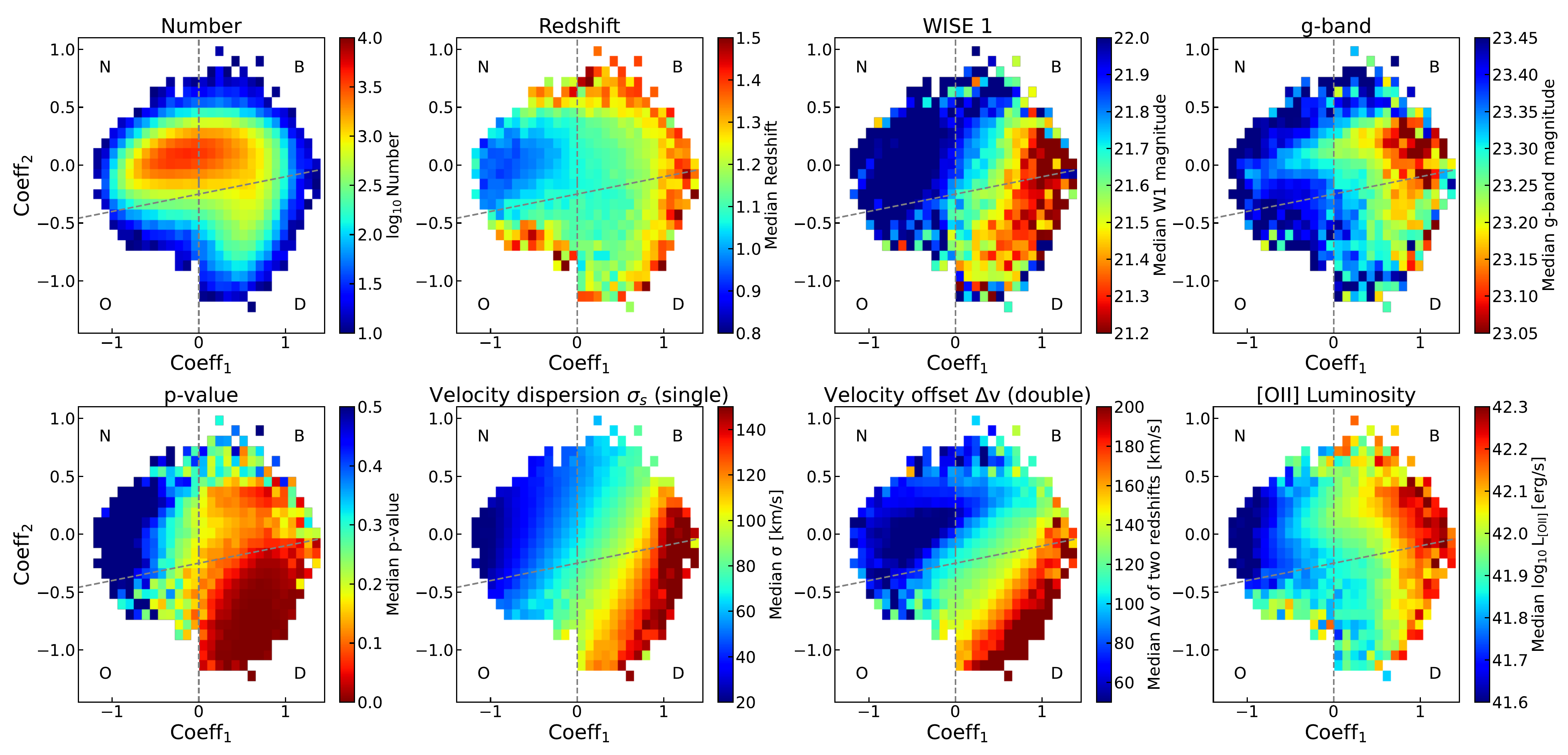}
\caption{Distributions of observed properties (upper panels) and spectral properties (lower panels) of ELGs as a function of $\rm Coeff_{1}$ (x-axis) and $\rm Coeff_{2}$ (y-axis). In the upper panels, the left panel shows the number of sources in each $\rm Coeff_{1}$ and $\rm Coeff_{2}$ bin. The 2nd to the 4th  panels show the median values of the redshifts, WISE 1 magnitude and g-band magnitude respectively. The lower panels from left to right show the median values of the p-value with the null hypothesis being that the $F_{d}$ model does not provide a better fit than the $F_{s}$ model, velocity dispersion $\sigma_{s}$, velocity offset ($\Delta v$), and [OII] luminosity respectively.
We note that for clarity, pixels with number of sources lower than 10, which occupy $\sim0.7\%$ of the total number of sources, are not displayed.
}
\label{fig:all_distribution}
\end{figure*}

\begin{figure*}
\center
\includegraphics[width=1\textwidth]{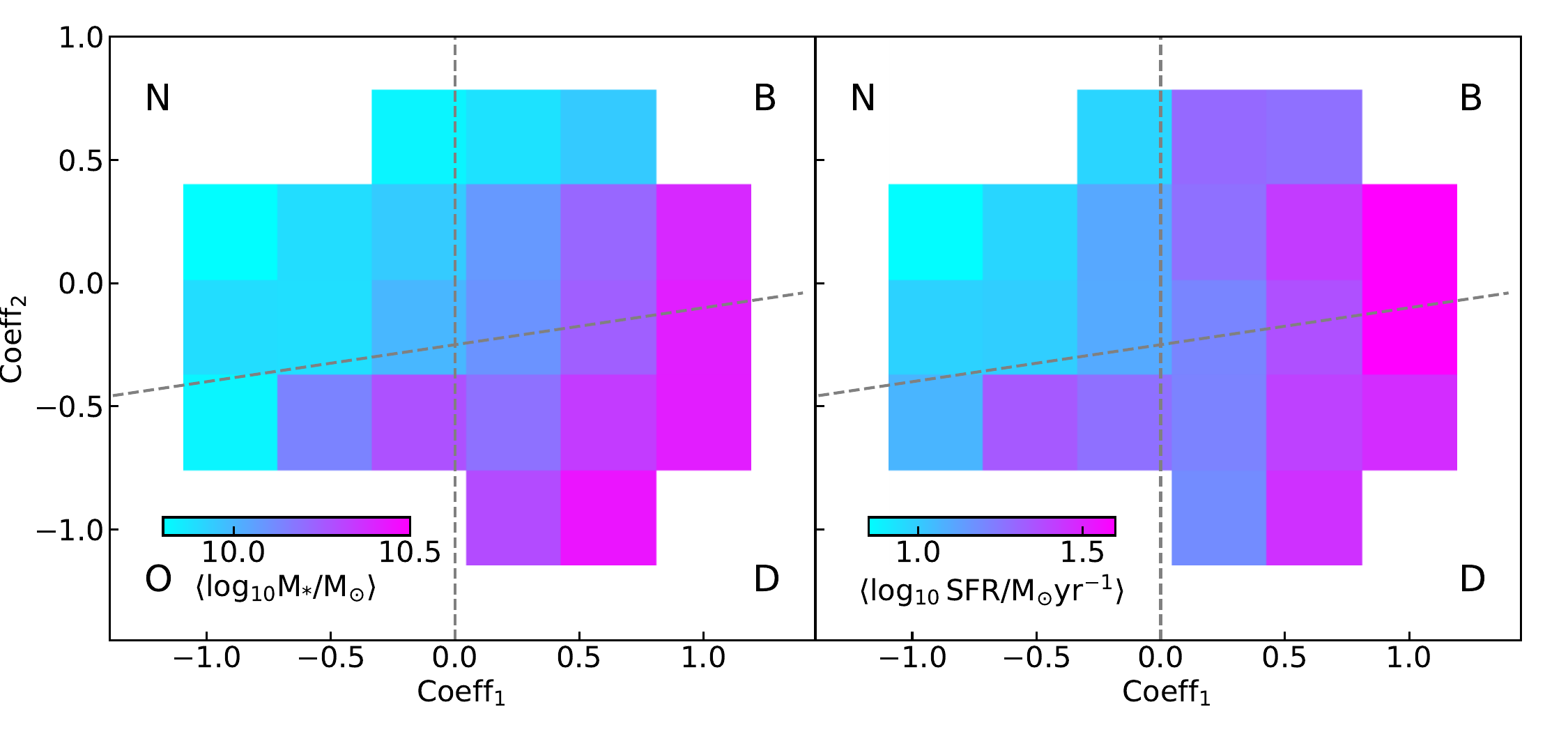}
\caption{
Stellar mass and SFR of ELGs as a function of $\rm Coeff_{1}$ (x-axis) and $\rm Coeff_{2}$ (y-axis). The colors in the left and right panels indicate the median stellar mass and SFR of ELGs respectively. 
For clarity, pixels with number of sources lower than 5, which occupy $\sim0.6\%$ of the total number of sources, are not displayed.
} 
\label{fig:sfr_mass_cosmos}
\end{figure*}

These four regions are highlighted with blue, green orange, and pink colors for narrow, broad, double-peak and outlier emission lines respectively in Figure~\ref{fig:coefficients}. The histograms on the top and right panels show the number distributions of $\rm Coeff_{1}$ and $\rm Coeff_{2}$ respectively. We find that $\sim 47\%$ of ELGs are in the narrow region, $\sim 38\%$ of ELGs are in the broad region, $\sim 13\%$ are in the double-peak region, and $\sim 2\%$ are in the outlier region. 
We classify the ELGs into four regions, 
and emphasize that while based on the coefficients, their distribution is smooth and continuous.

\subsection{Coefficients vs galaxy observed properties}
We now explore and describe the relationships between the observed properties of ELGs and $(\rm Coeff_{1}, Coeff_{2})$ values. In the upper panels of Figure~\ref{fig:all_distribution}, we first show the median values of galaxy observables in each $\rm Coeff_{1}$ and $\rm Coeff_{2}$ bin, including the number of ELGs, their redshifts, the observed \textit{WISE} 1 band magnitude, and the observed g-band magnitude from left to right respectively.

\textbf{Number:} The first upper panel from the left shows the number distribution of DESI ELGs as a function of $\rm Coeff_{1}$ and $\rm Coeff_{2}$. As shown in Figure~\ref{fig:coefficients}, most of the ELGs are in the narrow and broad regions with a small fraction of ELGs in the double-peak region.

\textbf{Redshifts:} The second upper panel shows the median redshifts as a function of $\rm Coeff_{1} - Coeff_{2}$. As can be seen, there is a redshift dependence between the median redshifts and $\rm Coeff_{1}$. ELGs with narrow \oii systems ($\rm Coeff_{1}<0$) have lower median redshifts than \oii with $\rm Coeff_{1}>0$. Additionally, ELGs with  $\rm Coeff_{1}$ and $\rm Coeff_{2}$ values near the edge of the entire distribution tend to have median redshifts $z\sim1.5$. This is
due to the fact that the line profiles of those galaxies are affected by sky emission line residuals and therefore have relatively extreme $\rm Coeff_{1}$ and $\rm Coeff_{2}$ values. 

\textbf{Observed WISE 1 band magnitude:} The third upper panel of Figure~\ref{fig:all_distribution} shows the median observed WISE 1-band magnitude of ELGs. There is a trend indicating that ELGs with larger $\rm Coeff_{1}$ values and lower $\rm Coeff_{2}$ values are on average brighter in WISE 1 band. 
This trend is not driven by the redshift correlation shown in the 2nd panel. It shows an opposite correlation that high-z ELGs with $\rm Coeff_{1}>0$ tend to have brighter observed magnitude than ELGs at low redshifts with $\rm Coeff_{1}<0$. 
The WISE 1 band ($3.4 \, \mu m$) corresponds to $\sim 1.7 \, \mu m$ in the rest-frame of ELGs at $z\sim1$, being sensitive to the stellar mass of galaxies. Therefore, the relationship between the coefficients and WISE 1 magnitude links the relationship between the coefficients and stellar mass.

\textbf{Observed g-band magnitude:} The last upper panel of Figure~\ref{fig:all_distribution} shows the median g-band observed magnitude. A trend can be observed, showing the brightness of ELGs in g-band increases with the $\rm Coeff_{1}$ values. ELGs with $\rm Coeff_{1}\sim1$ on average are brighter than the rest of the ELGs. Similarly to the WISE 1 band distribution, this trend is not driven by the redshift correlation shown in the 2nd panel. We also find that the relationship between the coefficients and g-band magnitude differs from the relationship between the coefficients and WISE 1 band magnitude. DESI ELGs tend to have bright median g-band magnitude around $\rm Coeff_{1}>0.5$ and $\rm Coeff_{2}\simeq0$, while DESI ELGs tend to have bright median WISE 1-band magnitude around $\rm Coeff_{1}>0.5$ but extends to $\rm Coeff_{2}\sim-1$. Considering the g-band covers the ultraviolet wavelength in the rest-frame of ELGs, this difference suggests that the relationship between the coefficients and SFR is not entirely driven by the relationship between the coefficients and stellar mass.

In addition to the photometric properties of DESI ELGs, we  extract the [OII] line information by performing spectral fitting analysis with two models.
The first one, $F_{s}$, is a single redshift model describing \oii lines with two Gaussian profiles with a single redshift,
\begin{equation}
    F_{s}(x;A_{s},\mu_{1},\sigma_{s}) = A_{s}\times e^{\frac{-(x-\mu_{1})}{2\sigma_{s}^{2}}} + 1.33 \, A_{s} \times e^{\frac{-(x-\mu_{1}-2.783)}{2\sigma_{s}^{2}}}, 
\end{equation}
where $x$ is the wavelength ($\rm \AA$), $A_{s}$ is the amplitude of the $\rm [OII]\lambda3726$ line, $\mu_{1}$ is the center wavelength  ($3727.092 \rm \AA$) of $\rm [OII]\lambda3726$ line, $\sigma_{s}$ is line width ($\rm \AA$) of the Gaussian profile. $\rm [OII]\lambda3729$ line is fitted by the second term where we adopt a fixed line ratio $\rm [OII]3729/[OII]3726=1.33$, the ratio between the two lines from the mean spectrum. 
The second model, $F_{d}$,  is a two redshift model describing \oii lines with two sets of two Gaussian profiles, $F_{s}$, with a wavelength offset $\Delta\lambda$, 
\begin{equation}
    \begin{aligned}
    & F_{d}(x; A_{d_{1}},\mu_{1},\sigma_{d_{1}},  R, \Delta\lambda,\sigma_{d_{2}}) = \\
    & F_{s}(x;A_{d_{1}},\mu_{1},\sigma_{d_{1}})+F_{s}(x;R \times A_{d_{1}},\mu_{1}+\Delta\lambda,\sigma_{d_{2}}),
    \end{aligned}
\end{equation}
where $R$ describes the amplitude difference in terms of ratio between the $\rm [OII]\lambda3726$ lines from two redshifts. 
With these two models, we obtain the best-fit parameters and the $\chi^{2}$ of the two models for all the ELG \oii profiles. The lower panels of Figure~\ref{fig:all_distribution} summarizes the fitting results.

\textbf{p-value:} 
We first quantify the performance of the two models for describing the \oii profiles with F-test by following \citet{Maschmann2020}. We estimate $f_{stat} = \frac{(\chi^2_{s}-\chi^2_{d})/(N_{s}-N_{d})}{\chi_{d}^{2}/N_{d}}$, where $\chi^2_{s}$ and $\chi^2_{d}$ are the $\chi^2$ values of the best-fits from the single and two redshift models and $N_{s}$ and $N_{d}$ are the corresponding degree of freedoms. We convert $f_{stat}$ values into the p-value for this hypothesis test with the null hypothesis being that the $F_{d}$ model does not provide a better fit than the $F_{s}$ model. In other words, lower p-values indicate a higher probability of rejecting the null hypothesis. 

The first lower panel of Figure~\ref{fig:all_distribution} from the left shows the median p-values, indicating that the p-values depend on the $\rm Coeff_{1}$ and $\rm Coeff_{2}$. For systems with negative $\rm Coeff_{1}$ values, the p-values are around 0.5, suggesting that the single redshift model is sufficient to describe the two [OII] lines. The median p-values decrease with the $\rm Coeff_{1}$ values.  This indicates that broader line profiles tend to require two components for describing the line profiles, especially the line profiles with three emission line peaks in the double-peak region which have median p-values lower than 0.05 (a commonly adopted threshold for rejecting the null hypothesis).

\textbf{Line velocity properties:} 
The second and the third lower panels of Figure~\ref{fig:all_distribution} show the median velocity dispersion of the emission lines estimated from the line width parameter $\sigma_{s}$ of the single redshift model and the median velocity separation estimated from the $\Delta \lambda$ parameter of the two redshift model respectively. 
We report the velocity dispersion with the spectral resolution effect being corrected. Both parameters increase from the top left corner ($\rm Coeff_{1}<0 \ \& \ Coeff_{2}>0$) to the bottom right corner ($\rm Coeff_{1}>0 \ \& \ Coeff_{2}<0$). The median velocity dispersion for DESI ELGs in the narrow region, broad region, the double-peak region are $\sim 50$ km/s, $\sim80$ km/s, and $\sim 100$ km/s respectively. 
For the systems that require two redshifts, velocity dispersion from the single redshift model partially reflects the velocity offset of the redshift difference shown in the third panel. The median velocity offsets for DESI ELGs in the broad and double-peak regions are $\sim 100$ km/s and $\sim 150$ km/s.

\textbf{[OII] luminosity:} 
Finally, the last lower panel of Figure~\ref{fig:all_distribution} shows the median total luminosity (erg/s) of the two [OII] lines estimated with the best-fit of the two redshift model. There is a correlation between [OII] luminosity and $\rm Coeff_{1}$, indicating that ELGs with broader line profiles have on average higher [OII] luminosity, a trend similarly to the g-band magnitude. This trend is detected across the entire redshift range probed in this work. Considering [OII] luminosity as a tracer of star-formation rate (SFR) of galaxies \citep[e.g.,][]{Kennicutt1998}, this result indicates that the SFR of ELGs correlates with the emission line profiles -- the broader the line, the higher the SFR. 

\begin{figure*}
\center
\includegraphics[width=1\textwidth]{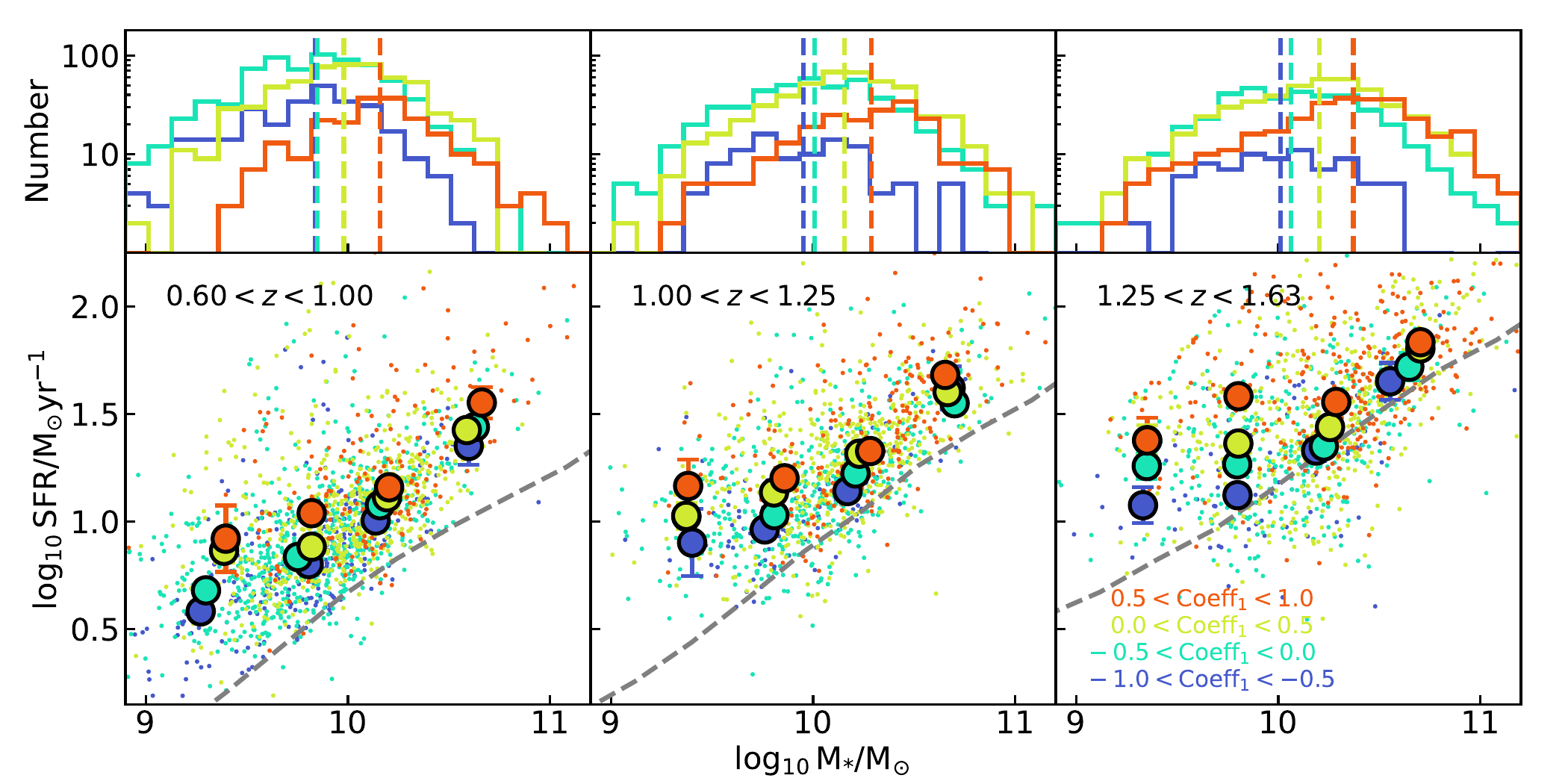}
\caption{Correlations between stellar mass, SFR, and $\rm Coeff_{1}$. The upper panels show the number distributions of ELGs with different $\rm Coeff_{1}$ values indicated by blue ($\rm -1<Coeff_{1}<-0.5$), cyan ($\rm -0.5<Coeff_{1}<0$), green ($\rm 0<Coeff_{1}<0.5$), and orange ($\rm 0.5<Coeff_{1}<1$). The vertical color dashed lines reflect the median stellar mass of the ELGs within the corresponding $\rm Coeff_{1}$ bin.
The three panels from left to right show the results for ELGs at three redshift bins from $z=0.6$ to $z=1.63$. The lower panels show the SFR as a function of stellar mass with colors indicating $\rm Coeff_{1}$. The small data points are individual ELGs and the large data points with uncertainty are the median SFR as a function of stellar mass for each $\rm Coeff_{1}$ bin. The uncertainties are estimated based on bootstrapping the sample 1000 times. The grey dashed lines in the lower panels are the median relationship based on galaxies in the COSMOS-SFG sample within the same redshift regions.} 
\label{fig:sfr_mass_coeff1}
\end{figure*}

\begin{figure}
\center
\includegraphics[width=0.45\textwidth]{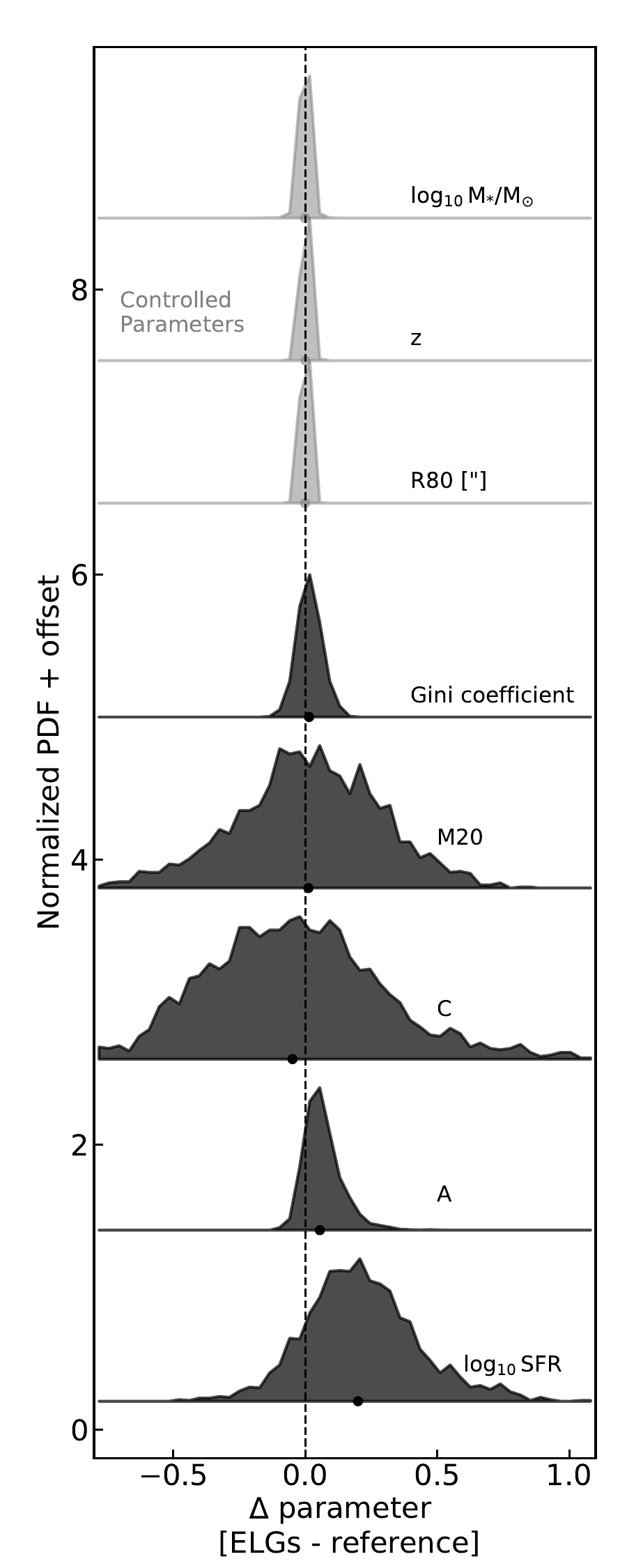}
\caption{Normalized distributions of the differences between the properties of DESI ELGs and that of the reference galaxies. The top three properties are $\rm log_{10}\, M_{*}/M_{\odot}$, redshift, and galaxy sizes (R80), which are controlled parameters. The lower five properties are Gini coefficients, M20, concentration C, galaxy asymmetry A, and $\rm log_{10}\, SFR$. Distributions are shifted in y-axis for clarity. The black data points indicate the median values of the distributions.} 
\label{fig:delta_pp}
\end{figure}

\begin{table*}[ht] 
\centering
\caption{Spearman's correlation coefficients between $\rm \Delta log_{10} SFR/M_{\odot}yr^{-1}$ and other parameters}
\begin{tabular}{c|ccccccc}
\hline  
& $\rm \Delta A$ & $\rm Coeff_{1}$ & $\rm \Delta M_{20}$ & $\rm Coeff_{2}$ & $\rm \Delta G$ & $\rm \Delta C$ & $\rm M_{*}$ \\
\hline
$\rm \Delta log_{10}\frac{SFR}{M_{\odot}yr^{-1}}$ & 
$0.31\pm0.02$ & 
$0.16\pm0.02$ & 
$0.12\pm0.02$ & 
$0.05\pm0.02$ & 
$0.05\pm0.02$ & 
$-0.09\pm0.02$ &
$-0.13\pm0.02$ \\
\hline
\end{tabular}
\label{table:sfr_other}
\end{table*}

\begin{table*}[ht] 
\caption{Spearman's correlation coefficients between PCA coefficients and morphological parameters}
\begin{tabular}{c|cccccc}
\hline  
& $\rm \Delta A$ & $\rm \Delta M_{20}$ & $\rm \Delta C$ & $\rm \Delta G$ \\
\hline
$\rm Coeff_{1}$ & $0.08\pm0.02$ & $0.02\pm0.02$ & $0.01\pm0.02$ & $0.00\pm0.02$ \\
\hline
$\rm Coeff_{2}$ & $0.07\pm0.02$ & $0.01\pm0.02$ & $0.05\pm0.02$ & $0.05\pm0.02$ \\
\hline
\end{tabular}
\label{table:pca_morh}
\end{table*}

\section{Relationships between line profiles and galaxy physical properties}

\subsection{Star-formation rate and stellar mass}
With the observed and spectral properties of DESI ELGs being explored, in order to better understand the underlying relationship, we now explore the physical properties of DESI ELGs and their connections to the \oii line profiles, using the DESI-COSMOS sample as described in Section 2.
Figure~\ref{fig:sfr_mass_cosmos} shows the median values of ELG stellar mass (left panel) and SFR (right panel) in the $\rm Coeff_{1}$-$\rm Coeff_{2}$ parameter space. The colors reflect the values of stellar mass and SFR. Being consistent with the trend observed in the median values of WISE 1 band magnitude and g-band magnitude shown in Figure~\ref{fig:all_distribution}, the stellar mass and SFR increase with $\rm Coeff_{1}$, demonstrating the overall relationship between the line profiles, stellar mass and SFR. 

Given that the stellar mass and SFR of star-forming galaxies correlate with each other, we further investigate the correlation between $\rm Coeff_{1}$ and  SFR with a fixed range of stellar mass of ELGs. The results are shown in Figure~\ref{fig:sfr_mass_coeff1} for three redshift bins from $0.6<z<1$ (left), $1<z<1.25$ (middle), and $1.25<z<1.63$ (right). The upper panels show the stellar mass distributions of ELGs with different $\rm Coeff_{1}$ ranges. The vertical color dashed line shows the median stellar mass for each $\rm Coeff_{1}$ bin, indicating that ELGs with higher $\rm Coeff_{1}$ values on average have higher stellar mass. 

The lower panels of Figure~\ref{fig:sfr_mass_coeff1} show the SFR of ELGs as a function of stellar mass with colors indicating their $\rm Coeff_{1}$ values. The data points with uncertainties are the median values of $\rm log_{10} SFR$ for the four $\rm Coeff_{1}$ bins. The uncertainties are estimated by bootstrapping the sample 1000 times. We find that with a fixed stellar mass, the SFR increases with $\rm Coeff_{1}$. This trend is observed across the stellar mass and redshift ranges of the DESI ELG sample. We also calculate the median SFR of the overall star-forming galaxy population at the same redshift range from the COSMOS-SFG sample defined in Sec 2.2.
The grey dashed line in each panel shows the corresponding median trend. 
As can be seen, the majority of DESI ELGs have SFR higher than the median SFR of the overall star-forming galaxy population. This result indicates that DESI preferentially selects star-forming galaxies with SFR higher than the main sequence galaxies, a conclusion also reached by \citet{Yuan2023} using DESI ELGs with $0.8<z<1.1$ based on a similar analysis.

To summarize, Figure~\ref{fig:sfr_mass_coeff1} demonstrates that while there is a general trend between the line profiles and the stellar mass and SFR which are coupled together due to the overall stellar mass and SFR correlation, there is an additional correlation between SFR and line profiles when considering ELGs with similar stellar mass. This trend is consistent with the correlation between gas velocity dispersion and SFR of star-forming galaxies observed at different redshifts from the local Universe \citep[e.g.,][]{Law2022}, $z\sim1$ \citep[e.g.,][]{Mai2024}, to $z\sim2$ \citep[e.g.,][]{Ubler2019}.
In the following, we explore the relationship between this excess SFR with respect to the overall star-forming galaxy population and lines profiles and investigate possible mechanisms behind this relationship.

\begin{figure*}
\center
\includegraphics[width=0.95\textwidth]{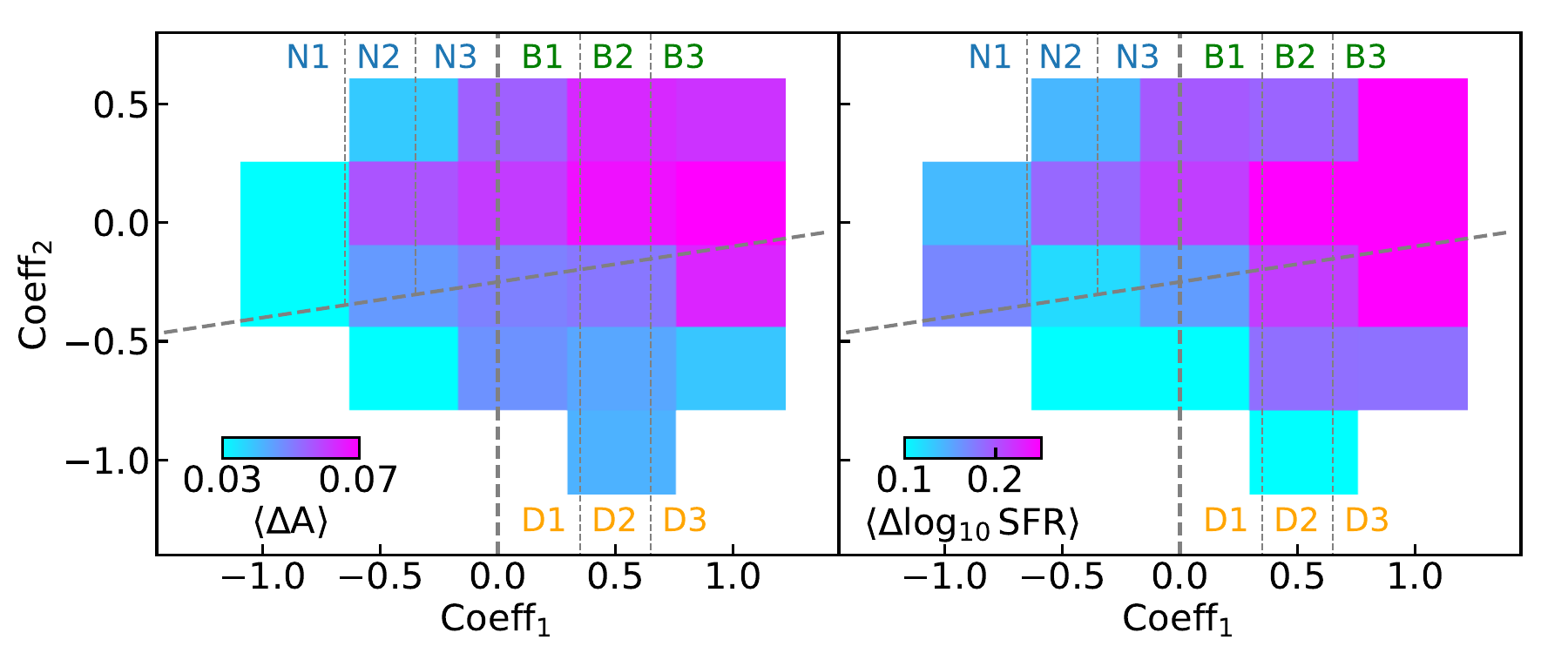}
\includegraphics[width=0.95\textwidth]{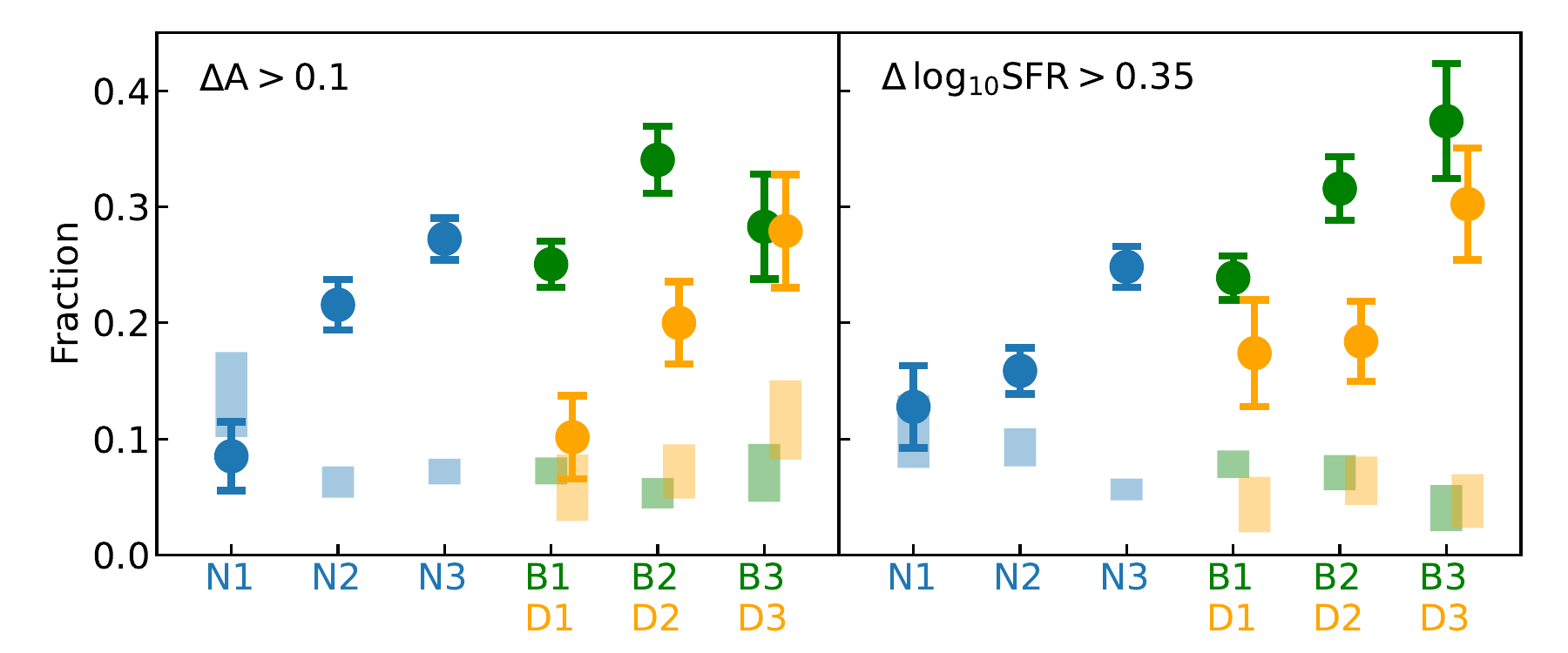}
\caption{Correlation between $\rm \Delta A$, $\rm \Delta log_{10} \, SFR$ and line profiles. The upper panels show the median $\rm \Delta A$ (left) and $\rm \Delta log_{10}\,  SFR/M_{\odot}yr^{-1}$ (right) as a function of $\rm Coeff_{1}$ and $\rm Coeff_{2}$.
The lower panels show the fractions of ELGs with $\rm \Delta A>0.1$ (left) and $\rm \Delta log_{10}\, SFR>0.35$ (right) within 9 $\rm Coeff_{1}-Coeff_{2}$ regions from narrow (blue), broad (green), to double-peak (orange) respectively. 
The color boxes show the same measurements for a control sample consisting of star-forming galaxies selected to have similar stellar mass, redshifts, and sizes as the DESI ELGs. The uncertainties are based on bootstrapping the sample 1000 times.
} 
\label{fig:Coeff_12_median}
\end{figure*}

\subsection{Excess SFR, galaxy morphology, and line profiles}
To explore how the excess SFR of ELGs correlates with the line profiles and inform the underlying mechanisms, in addition to the stellar mass and SFR, we use the DESI-ZEST sample which includes five morphological properties derived from HST ACS F814W images from the ZEST catalog \citep{COSMOS_zbest} in the analysis: asymmetry (A), Gini coefficient (G), concentration (C), and the second-order moment of the brightest $20\%$ pixels (M20), and R80 parameter reflecting the galaxy size of ELGs. In Appendix~\ref{appendix:morphology}, we summarize the average relationships between line profiles and the morphological properties.

We calculate the differences between the physical and morphological parameters of DESI ELGs and that of overall star-forming galaxies from the COSMOS-ZEST sample at the similar redshifts, stellar mass, and sizes.
This analysis allows us to remove the correlations driven by these three parameters. More specifically, for each DESI ELG, we use at least 10 COSMOS-ZEST galaxies as references with redshift difference $|\Delta z|<0.025$, stellar mass difference $\rm |\Delta log_{10} \, M_{*}/M_{\odot}|<0.05$, and size difference $\rm |\Delta R80|<0.025"$. If less than 10 COSMOS-ZEST galaxies are found, we increase the conditions until there are at least 10 galaxies within the selection. Finally, for each ELG, we calculate the difference between the ELG parameter values, SFR, A, G, M20, and C, and the median values of their references as $\Delta \rm log_{10}\, SFR$, $\Delta \rm A$, $\Delta \rm G$, $\Delta \rm M20$, $\Delta \rm C$ respectively.

Figure~\ref{fig:delta_pp} shows the normalized probability density distributions of the excess values of the parameters. The top three parameters are the controlled parameters, stellar mass, redshift, and size (R80). By construction, the distributions are narrow and center around 0. The lower 5 parameters are the morphological parameters, G, M20, C, A, and the SFR values. We find that while there are mild positive excess of G and M20 parameters and on average lower C parameter in comparison to the reference galaxies, the most significant difference is the asymmetric parameter A. The majority ($\sim85\%$) of ELGs have the asymmetric parameter A values higher than the reference galaxies with $\rm \Delta A>0$, a trend similar to the trend of the SFR with $\sim86\%$ ELGs with $\Delta \rm log_{10}\, SFR>0$.

To first explore the correlations between parameters, Table~\ref{table:sfr_other} shows the Spearman's correlation coefficients ($\rho_{s}$) between $\Delta \rm log_{10}\, SFR$ and the spectral PCA coefficient 1 and 2 and the excess values of morphological parameters with the uncertainties estimated by bootstrapping the sample 1000 times. We order the parameters according to their values of Spearman's correlation coefficients. The results show that
\begin{itemize}
    \item the  $\Delta \rm log_{10}\, SFR$ correlates with $\rm \Delta A$ with the highest absolute $\rho_{s}\simeq0.31\pm0.02$ value among the parameters;

    \item the second parameter with high $\rho_{s}\simeq0.16\pm0.02$ value is the line profile coefficient, $\rm Coeff_{1}$. We note that $\rho_{s}$ between $M_{*}$ and $\rm \Delta log_{10} SFR$ is $-0.13\pm0.02$. This demonstrates that the correlation between $\rm Coeff_{1}$ and $\Delta \rm log_{10}\, SFR$ is not driven by stellar mass. 
    
    \item $\Delta M_{20}$, $\rm Coeff_{2}$, and $\Delta C$ parameters also have some correlations with  $\Delta \rm log_{10}\, SFR$. 
\end{itemize}
These results demonstrate that $\Delta \rm log_{10}\, SFR$ of DESI ELGs correlates with both the morphological structures and the spectral profiles of galaxies. 

To better understand the interplay between the shape of galaxies and the line profiles, Table~\ref{table:pca_morh} shows the $\rho_{s}$ between $\rm Coeff_{1}$ and $\rm Coeff_{2}$ and the excess values of morphological parameters with the uncertainties estimated by bootstrapping the sample 1000 times. As can be seen,  $\rm Coeff_{1}$ and $\rm Coeff_{2}$ correlate the most with the $\rm \Delta A$ ($> 3 \sigma$ detection) and $\rho_{s}$ with other morphological parameters are consistent with no detection ($< 3 \sigma$). In other words, based on the $\rho_{s}$ values of different parameter pairs, we find that there is a relationship between $\rm \Delta log_{10} SFR$, $\rm \Delta A$, $\rm Coeff_{1}$, and $\rm Coeff_{2}$. 

We estimate the median $\rm \Delta A$ and $\rm \Delta log_{10} SFR$ as a function of $\rm Coeff_{1}$ and $\rm Coeff_{2}$ to quantify the relationship. The results are shown in the upper panels of Figure~\ref{fig:Coeff_12_median}, indicating that both median $\rm \Delta A$ (left panel) and median $\rm \Delta log_{10} SFR$ (right panel) increase from negative $\rm Coeff_{1}$ to positive $\rm Coeff_{1}$ and from negative $\rm Coeff_{2}$ to positive $\rm Coeff_{2}$. 

We also separate the $\rm Coeff_{1}-Coeff_{2}$ parameter space into 9 regions based on general classification shown in Figure~\ref{fig:coefficients}: 
\begin{itemize}
    \item Narrow region: N1 ($\rm Coeff_{1}<-0.65$), N2 ($\rm -0.65<Coeff_{1}<-0.35$), and N3 ($\rm -0.35<Coeff_{1}<0$);
    \item Broad region: B1 ($\rm 0<Coeff_{1}<0.35$), B2 ($\rm 0.35<Coeff_{1}<0.65$), B3 ($\rm Coeff_{1}>0.65$);
    \item Double-peak region: D1 ($\rm 0<Coeff_{1}<0.35$), D2 ($\rm 0.35<Coeff_{1}<0.65$), D3 ($\rm Coeff_{1}>0.65$).
\end{itemize}
We calculate the fraction of ELGs with high $\rm \Delta A$ and $\rm \Delta log_{10} SFR$ in each region. The lower panels of Figure~\ref{fig:Coeff_12_median} summarize the results. The lower left panel show the fractions for $\rm \Delta A>0.1$. We find that the fractions of ELGs with high $\rm \Delta A$ increase from the narrow region ($\rm Coeff_{1}<0$) to the broad region ($\rm Coeff_{1}>0$). The fractions can increase by a factor of 3 from $\sim0.1$ to $\sim 0.3$. Moreover, for ELGs with $\rm Coeff_{1}>0$, the fraction is higher for ELGs in the broad region ($\rm Coeff_{2}\sim0$) than in the double-peak region ($\rm Coeff_{2}\leq-0.4$).
The lower right panel shows the same measurements with $\rm \Delta log_{10} SFR>0.35$. The overall trends of $\rm \Delta log_{10} SFR$ are consistent with $\rm \Delta A$, showing that ELGs with line profiles in the broad region (B) have relatively higher asymmetry morphology and higher SFR than their reference star-forming galaxies with similar stellar mass, sizes and at the same redshifts. 
In addition, we perform the same analysis for a control sample which consists of star-forming galaxies from the COSMOS-ZEST sample selected to have similar stellar mass, redshifts, and sizes as the properties of DESI ELGs with $|\rm \Delta log_{10} M_{*}/M_{\odot}|<0.05$, $|\Delta z|<0.025$ and $|\Delta R80|<0.025$ respectively. The color boxes in the lower panels of Figure~\ref{fig:Coeff_12_median} show the results, indicating that the DESI ELGs have preferentially higher $\rm \Delta A$ and $\rm \Delta log_{10} SFR$ than typical star-forming galaxies.

\begin{figure}
\center
\includegraphics[width=0.49\textwidth]{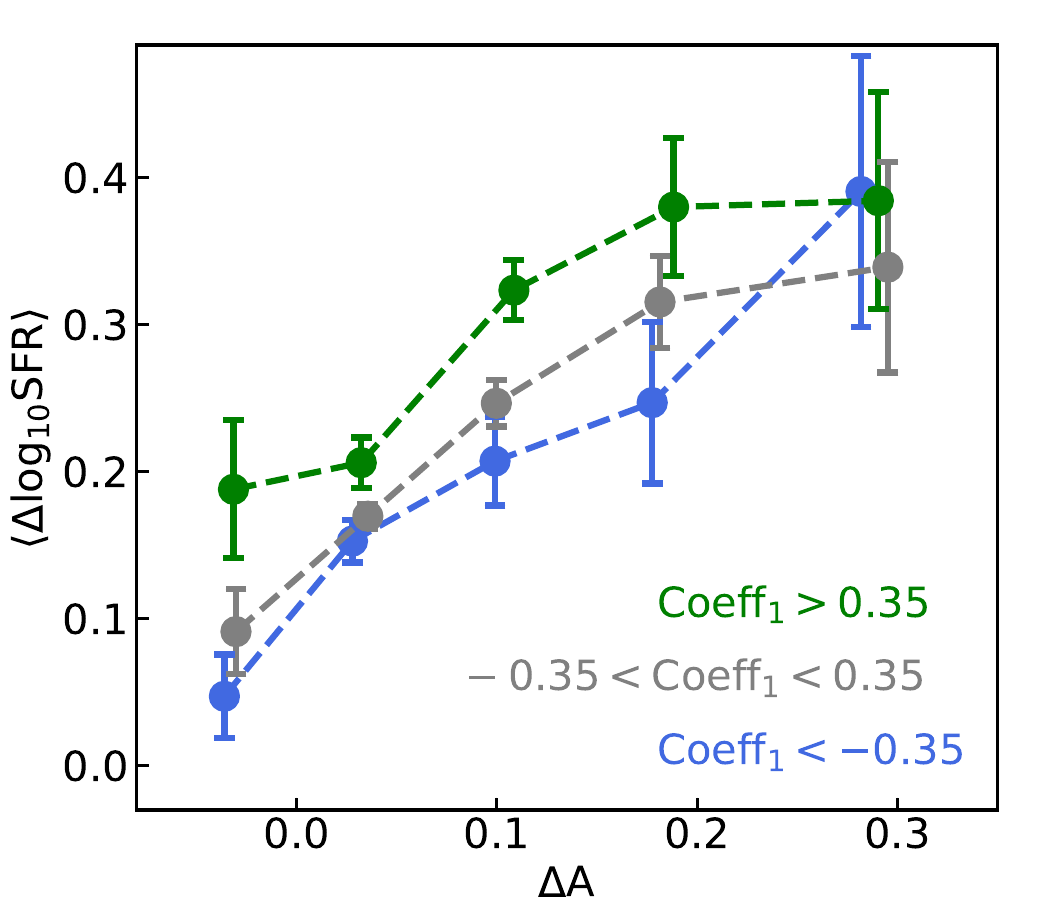}
\caption{Median $\rm \Delta log_{10}\, SFR$ as a function of $\rm \Delta A$ and $\rm Coeff_{1}$. The blue, grey, and green data points show the median $\rm \Delta log_{10}\, SFR$ as a function of $\rm \Delta A$ for ELGs with $\rm Coeff_{1}<-0.35$, $\rm -0.35<Coeff_{1}<0.35$, and $\rm Coeff_{1}>0.35$ respectively. These measurements only include the ELGs in the narrow and broad regions with uncertainties based on bootstrapping the sample 1000 times.} 
\label{fig:aa_sfr_coeff1_2}
\end{figure}

Finally, we quantify the median $\rm \Delta log_{10}\, SFR$ as a function of $\rm \Delta A$ and line profiles with the results shown in Figure~\ref{fig:aa_sfr_coeff1_2}. The blue and green data points are measurements from ELGs with $\rm Coeff_{1}<-0.35$ and $\rm Coeff_{1}>0.35$ respectively. The grey data points show the measurements in between. Here we only include ELGs within the narrow and broad regions with $\rm Coeff_{2}>0.15 \times Coeff_{1}-0.25$. The measurements indicate that there is a general correlation between $\rm \Delta log_{10}\, SFR$ and $\rm \Delta A$ which has the same slope but different offsets for galaxies with different $\rm Coeff_{1}$. With similar $\rm \Delta A$ values, DESI ELGs with broader [OII] lines have, on average, 0.1 dex higher $\rm \Delta log_{10} SFR$ than those with narrow [OII] lines. This again illustrates the connection between [OII] line profiles, $\rm \Delta A$ and $\rm \Delta log_{10} SFR$ of DESI ELGs.

\begin{figure*}
\center
\includegraphics[width=1.\textwidth]{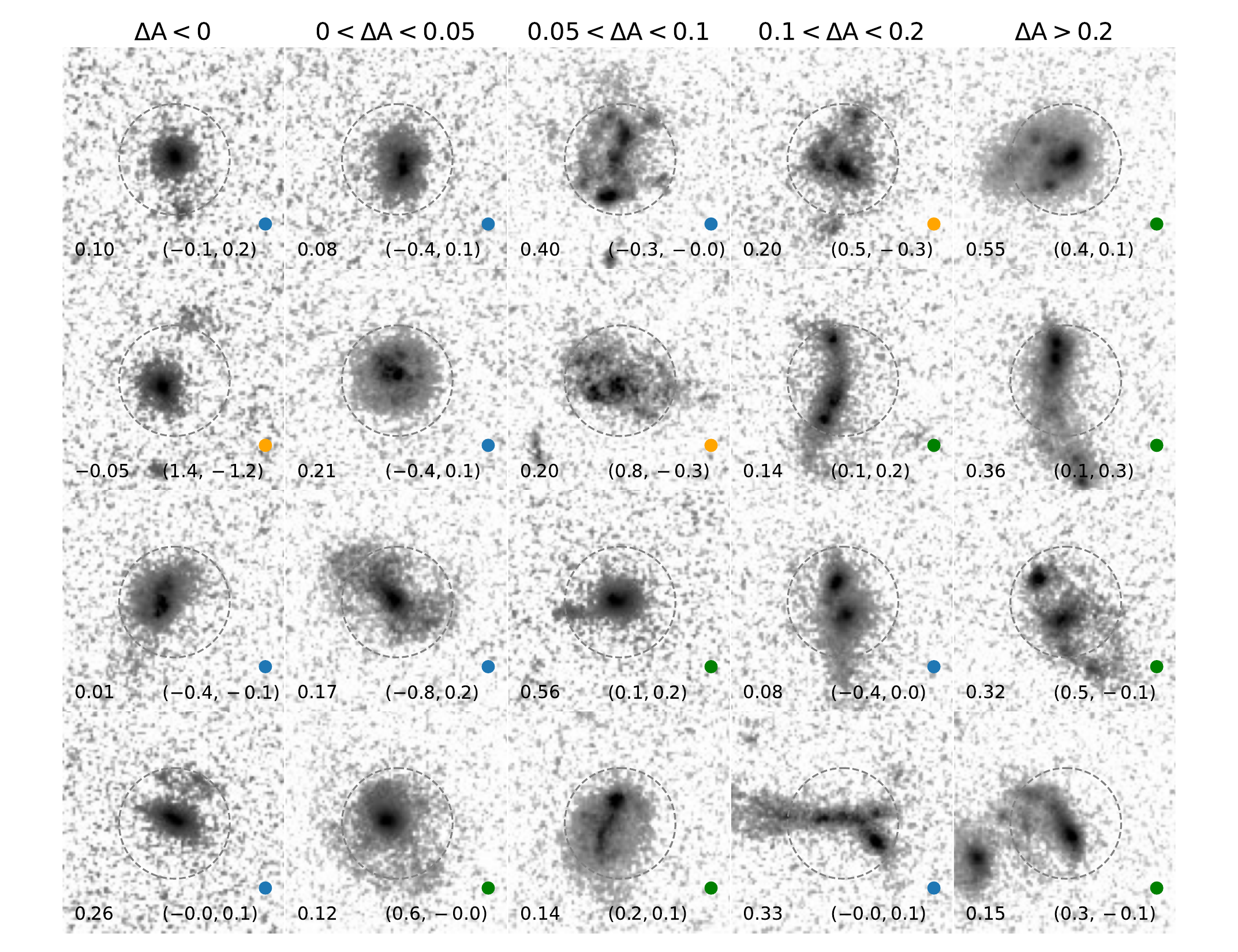}
\caption{Example images ($3"\times3"$) of DESI ELGs with different $\rm \Delta A$ values. Each column shows the HST ACS F814W images of four randomly selected DESI ELGs within the $\rm \Delta A$ selection as shown on the top of the column. The $\rm \Delta A$ values increase from left to right respectively. The values listed on the lower-left corner of the images are the $\rm \Delta log_{10} \, SFR$ of the ELGs and the values on the right are $\rm (Coeff_{1},Coeff_{2})$. 
The color dots on the lower-right corner indicate the spectral regions of the ELGs with blue, green, and orange for narrow, broad, and double-peak regions respectively.
The grey circle indicates the size of the DESI fiber ($1.5"$ in diameter).} 
\label{fig:images}
\end{figure*}

\begin{figure}
\center
\includegraphics[width=0.4\textwidth]{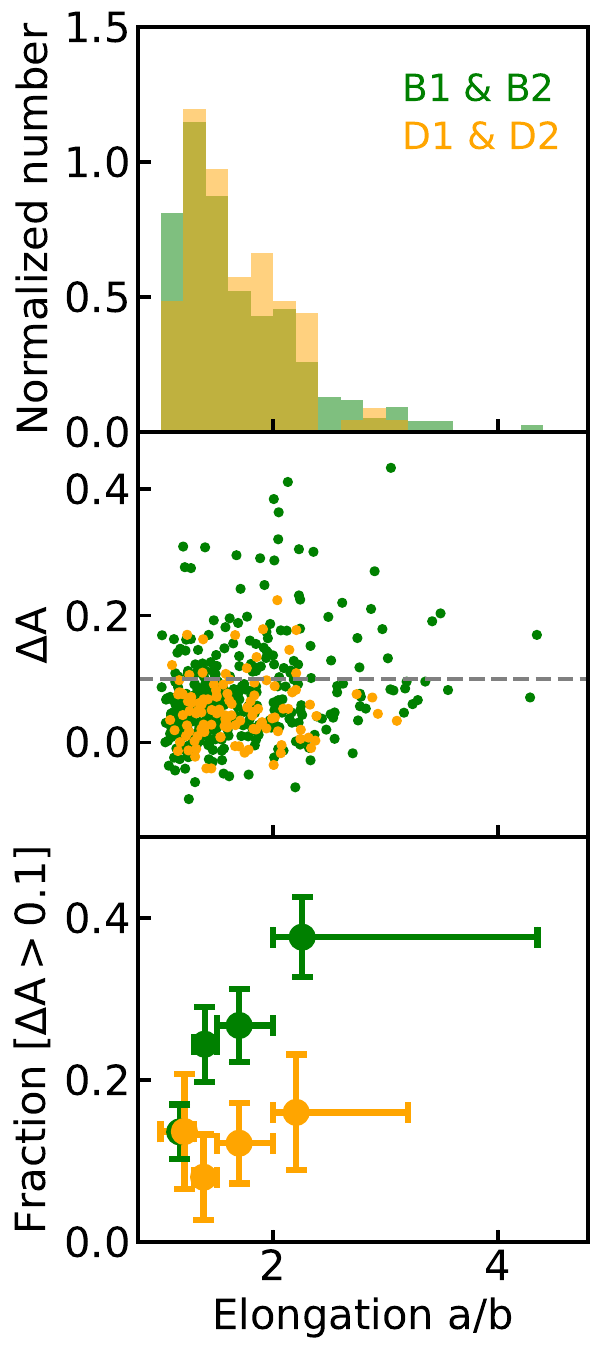}
\caption{Relation between $\rm \Delta A$ and galaxy elongation of ELGs with $\rm 10<log_{10}\,M_{*}/M_{\odot}<10.5$. \textit{Top:} Normalized number distributions of ELGs in $\rm B1 \, \&\, B2$ regions (green) and in $\rm D1 \, \&\, D2$ regions (orange) as a function of galaxy elongation. \textit{Middle:} $\rm \Delta A$ distribution as a function of galaxy elongation for the two ELG samples. \textit{Bottom:} Fractions of ELGs with $\rm \Delta A>0.1$ as a function of galaxy elongation for the two ELG samples. The uncertainties are estimated by bootstrapping the sample 1000 times. } 
\label{fig:AB_only}
\end{figure}

\section{Discussion}
\subsection{The physical mechanisms driving the relationship between SFR, asymmetry, and line profiles}

Utilizing data from the DESI spectroscopic observations and the COSMOS multi-wavelength deep images, we find a relationship between three parameters of DESI ELGs, excess SFR ($\rm \Delta log_{10} SFR$), the asymmetry of galaxy shape ($\rm \Delta A$), and the [OII] line profiles. In the following, we discuss two scenarios that can possibly explain the observed relationship. 

\textbf{Merger scenario:} 
The merging of galaxies has been considered as one of the crucial processes that can transform star-forming galaxies into passive ones. Theoretical works have shown that when two gas-rich galaxies merge, gas flows into the centers of the galaxies and trigger excess star-formation activities \citep[e.g.,][]{Hopkins2008,Lotz2008,Patton2020,Bottrell2023}. This merger-induced star-formation activities have also been supported by observational studies. Various observational approaches have been used to identify merging galaxies. For example, making use of data from the Sloan Digital Sky Surveys, one can identify galaxy pairs with a range of impact parameters and use such a sample to explore the difference of the SFR compared with the SFR of controlled isolated galaxies with similar mass \citep[e.g.][]{Ellison2008, Ellison2010, Behroozi2015}. Another approach is to use the morphological parameters of galaxies derived from imaging datasets to identify possible shape features induced by merging events, such as tidal structures and/or asymmetry \citep[e.g.,][]{Lotz2008,Yesuf2021}. 
Both simulations and observations have shown that merging events at the closest separation enhance the SFR by approximately a factor of 2 with a fixed stellar mass. 

Our results can be explained by this galaxy merger scenario. The excess asymmetry value $\rm \Delta A$ indicates that a sizable fraction of DESI ELGs are disturbed galaxies, e.g.\ there are approximately $25\%$ of DESI ELGs with $\rm \Delta A>0.1$, 
while there are only $7\%$ of the controlled sample with $\rm \Delta A>0.1$
(lower panels of Figure~\ref{fig:Coeff_12_median}).
Figure~\ref{fig:images} shows the HST ACS F814W images of randomly selected DESI ELGs as a function of $\rm \Delta A$. For most of the galaxies with $\rm \Delta A>0.05$, two or multi clumpy structures can be observed at small scales ($<1"$) with disturbed and possibly tidal features. 
The correlation between $\rm Coeff_{1}$ and $\rm \Delta A$ values provide a further support on this merger scenario --- when two interacting galaxies are at a close distance, the emission lines of the two galaxies are included in a single DESI fiber (12 kpc in diameter at $z\sim1$) with a moderate line-of-sight velocity difference $\Delta v$ which will produce a \oii line profile broader than a typical isolated star-forming galaxy with similar stellar mass. The three observational properties of DESI ELGs, the line profile, the asymmetry of galaxy shape, and the excess SFR, can be all fit within the merger scenario simultaneously.

\textbf{Disk instability scenario:} Another possibility is that the multi clumpy structures shown in Figure~\ref{fig:images} are star-forming regions formed due to disk instability \citep[e.g.,][]{Dekel2009, Mandelker2014} in a single galaxy. Previous studies \citep[e.g.,][]{Guo2015, Murata2014, Martin2023, Sattari2023} have shown that the clumpy fraction, the ratio between the number of star-forming galaxies with at least one off-center clump and the total number of star-forming galaxies, roughly peaks between $z\sim1$ and $z\sim2$, overlapping the redshift region of the DESI ELGs, and correlates with SFR with a fixed stellar mass. 
Bright star-forming clumpy structures can lead to asymmetric light distribution with possible associated emission lines \citep[e.g.,][]{Fisher2017, Altamirano2018} that contribute to the emission line profiles. 
Moreover, the disk instability is expected to drive turbulent which in turn enhances gas velocity dispersion in galaxies \citep[e.g.,][]{Goldbaum2016}.
Therefore, this scenario can also produce correlations between $\rm \Delta SFR$, $\Delta A$, and line profiles as observed for DESI ELGs. 

The origins of clumpy structures observed in high-redshift galaxies are still under active investigations. Some studies have shown that the clump size and gas kinematics relationship of some clumpy galaxies is consistent with the prediction of the disk instability scenario \citep[e.g.,][]{Fisher2017}. However, \citet[][]{Ribeiro2017} have shown that galaxies with only two major clumps tend to have clump mass being inconsistent with the expected mass based on disk instability. The authors argue that galaxies with two major clumps can be ongoing galaxy mergers. In addition, \citet[][]{Elmegreen2021} have shown that the observed appearance of major mergers at high redshifts is similar to the observed appearance of clumpy isolated disk galaxies. These studies indicate the complexity of distinguishing clumps formed by disk instability and galaxy mergers. 
In the DESI ELG sample, while there are galaxies having only two major clumps, some galaxies have more than two clumps as can be seen in Figure~\ref{fig:images}. Therefore, it is possible that DESI ELGs include both types of systems. 

While estimating the contributions of these two scenarios in the overall DESI ELG population is beyond of the scope of this work, one can combine multi-wavelength deep images with radio observations for spatially resolved gas properties to identify tidal features, characterize the properties of the clumpy structures and resolve the nature of these galaxies.

\subsection{The origins of double-peak galaxies}
Our results show that while ELGs in the double-peak region tend to have higher velocity offsets between two velocity components than ELGs in the broad region, ELGs in the double-peak region, especially in D1 and D2 regions, have on average lower $\rm \Delta A$ and $\rm \Delta log_{10}\, SFR$ than ELGs located in the broad (B) region with $\rm Coeff_{1}>0$ and $\rm Coeff_{2}\simeq 0$, as shown in Figure~\ref{fig:Coeff_12_median}.

We propose that this difference might be due to the fact that instead of tracing two galaxies during the merging process or galaxies with violent disk instability, there is higher fraction of ELGs in the double-peak region with the velocity components reflecting the rotating disks of the galaxies without undergoing SFR enhancement events. One possible way to explore these two scenarios is to investigate the galaxy inclination and $\rm \Delta A$ relation of ELGs. The expectation for the rotating disk scenario is that those galaxies tend to be more edge-on with low $\rm \Delta A$. On the other hand, the morphology of two merging galaxies with close separation or galaxies with two major clumps can resemble the morphology of edge-on galaxies. Therefore, merging/clumpy galaxies also tend to look like edge-on galaxies but with higher $\rm \Delta A$ than the rotating disks. 

In Figure~\ref{fig:AB_only}, we examine the distribution of $\rm \Delta A$ and galaxy inclination, using the elongation parameter, the ratio between the lengths of semi-major and semi-minor axes of galaxies, from HST images as a proxy \citep{Leauthaud2007}. We only select ELGs with $\rm 10<log_{10}M_{*}/M_{\odot}<10.5$ to reduce possible effects associated with galaxy mass. The top panel shows the normalized number distributions of the elongation parameter of DESI ELGs in $\rm B1\, \& \, B2$ regions and $\rm D1\, \& \, D2$ regions indicated by green and orange respectively. The middle panel shows the distribution of $\rm \Delta A$ as a function of the elongation parameter and the bottom panel shows the fraction of ELGs with $\rm \Delta A>0.1$ as a function of the elongation parameter for the two types of ELGs. We find that ELGs in  $\rm B1\, \& \, B2$ regions and $\rm D1\, \& \, D2$ regions both have a broad range of elongation parameter. The middle and bottom panels show that for ELGs in $\rm D1\, \& \, D2$ regions, the fraction of sources with $\rm \Delta A>0.1$ is consistently being $\sim 0.1$ across all elongation parameter values. In contrast, the fraction of ELGs in $\rm B1\, \& \, B2$ regions with $\rm \Delta A>0.1$ increases with the elongation parameter from 0.2 to 0.4. These observed trends are consistent with the above proposed scenario that mergers/clumpy galaxies tend to have high elongation and high $\rm \Delta A$, while rotating disks tend to have high elongation but with low $\rm \Delta A$. However, we note that the above results are suggestive and spatially resolved kinematics information is needed to conclusively identify the origins of double-peak emission lines observed in DESI ELGs. 

Previous studies have also suggested that double-peak emission lines observed in galaxies are associated with rotating disks at the local Universe. 
For example, \citet{Maschmann2020} found $\sim 5000$ double-peak emission line galaxies from the SDSS spectroscopic dataset. Based on simulations, \citet{Maschmann2023} concluded that the double-peak emission lines can be originated from the central rotating bars of galaxies or minor mergers. \citet{Chen2016} also found that rotating galaxy discs can explain the double-peak emission lines observed in SDSS disc star-forming galaxies.

\subsection{DESI ELG selection for galaxy properties}

\begin{figure*}
\center
\includegraphics[width=0.9\textwidth]{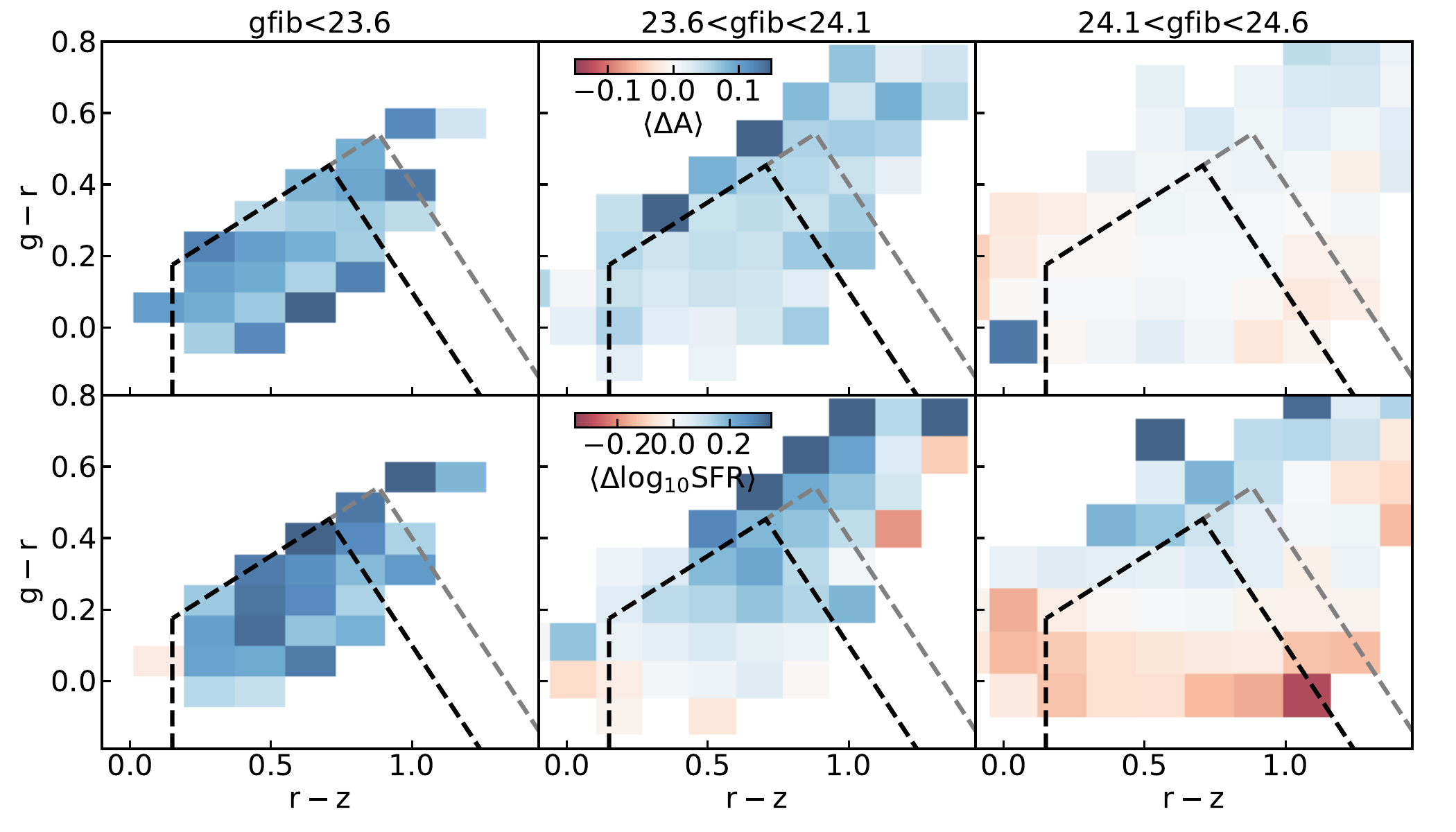}
\caption{Median $\rm \Delta A$ values and $\rm \Delta log_{10} \, SFR$ of star-forming galaxies as a function of $g-r$, $r-z$, and g-band fiber magnitude ($\rm gfib$). Galaxies are selected with $\rm log_{10} M_{*}/M_{\odot}>9$, $\rm 1.1<z_{photo}<1.63$ and $g>20$ from COSMOS2020 catalog with HST ACS morphology measurements. The upper panels are median $\Delta A$ values as a function of galaxy colors with $\rm gfib<23.6$ (left), $\rm 23.6<gfib<24.1$ (middle) and $\rm 24.1<gfib<24.6$ (right). The lower panels are median $\Delta log_{10} \, SFR$ as a function of galaxy colors with $\rm gfib<23.6$ (left), $\rm 23.6<gfib<24.1$ (middle) and $\rm 24.1<gfib<24.6$ (right).} 
\label{fig:DESI_selection}
\end{figure*}

We now explore the key selection criteria for DESI ELGs which preferentially include galaxies with high SFR and asymmetric morphology. The DESI ELG selections for the main survey are summarized in Table 2 of \citet{DESI_ELG}, which includes 
\begin{itemize}
    \item $g>20$, 
    \item $\rm g \ band\ fiber\ magnitude \ (gfib)\  <24.1$, and 
    \item $g-r$ and $r-z$ color boundaries. 
\end{itemize}
To investigate how these selection conditions affect the galaxy properties, we use the COSMOS-ZEST galaxy sample and cross-match the sample with the photometric catalog\footnote{DR10 \url{https://www.legacysurvey.org/dr10/catalogs/}} from the DESI Legacy Imaging Surveys \citep{Dey2019} to obtain the photometric information used in the DESI target selections. 
We then select star-forming galaxies with $\rm log_{10} M_{*}/M_{\odot}>9$, $\rm 1.1<z_{photo}<1.63$ (the parameter space that covers the main DESI ELGs) and $g>20$.
For each star-forming galaxy, we follow the same procedure for estimating $\rm \Delta A$ and $\rm \Delta log_{10} SFR$ as the DESI ELG sample by searching for at least 10 galaxies as references with redshift difference $|\Delta z|<0.025$, stellar mass difference $\rm |\Delta log_{10} \, M_{*}/M_{\odot}|<0.05$, and size difference $\rm |\Delta R80|<0.025"$ and increasing the conditions if less than 10 galaxies are found. The $\rm \Delta A$ and $\rm \Delta log_{10} SFR$ are the differences between the values of the galaxies and the median values of the reference galaxies. 

With $\rm \Delta A$ and $\rm \Delta log_{10} SFR$ information, we 
calculate the median $\rm \Delta A$ and $\rm \Delta log_{10} SFR$ as a function of $g-r$, $r-z$ colors, and $\rm gfib$. 
The upper panels of Figure~\ref{fig:DESI_selection} show the median $\rm \Delta A$ values across the $g-r$ and $r-z$ color space with $\rm gfib<23.6$ (left panel), $\rm 23.6<gfib<24.1$ (middle panel) and $\rm 24.1<gfib<24.6$ (right panel). The black dashed lines are the boundaries for the DESI ELGs LOP sample and the grey dashed lines are the extended region for the VLO sample. 
As can be seen, galaxies that are brighter in $\rm gfib$ preferentially have higher $\rm \Delta A$ values.
A similar trend is observed for $\rm \Delta log_{10} SFR$ as shown in the lower panels of Figure~\ref{fig:DESI_selection} indicating the median $\rm \Delta log_{10} SFR$ values of galaxies. 
We note that while there is no strong correlation between $\rm \Delta A$ and galaxy colors, a correlation between $\rm \Delta log_{10} SFR$ and galaxy colors can be observed in the middle and right panels. 
Therefore, we conclude that the g-band fiber magnitude selection is the primary factor for selecting galaxies with more disturbed morphology and higher SFR than the overall star-forming galaxy population at the same redshifts and the (g-r, r-z) color selections are an additional factor modulating the SFR. 
Note that while $\rm gfib\sim24.1$ reaches the limiting magnitude of the Legacy Surveys \citep{Dey2019} with larger uncertainties, we have performed the same analysis with g, r, z-band images with depths $\sim27-28$ magnitude from Hyper Suprime-Cam \citep{Aihara2018, Aihara2019} in the COSMOS2020 catalog which yield consistent results.

\subsection{Implications for DESI ELG clustering measurements}

Recent studies have shown that in order to reproduce the DESI ELG clustering properties, especially the small-scale signals ($\rm \leq 0.2 \, Mpc \, h^{-1}$), and obtain physically motivated ELG-halo connection models, additional parameters are required in the halo occupation distribution (HOD) modeling \citep[e.g.,][]{Gao2023, Yuan2023, Rocher2023, Gao2024}. These results indicate that the abundance of satellite ELGs in the halos depends on the properties of the central ELGs --- a phenomenon called "conformity" \citep[][for a review]{Wechsler2018}.

By exploring the properties of ELGs at $0.8<z<1.1$ and their morphology with COSMOS dataset, \citet{Yuan2023} postulate that such conformity signals are driven by galaxy merging induced star formation in both central and satellite galaxies. Our findings of the relationship between SFR, galaxy morphology and the line profiles are consistent with this scenario. In addition, the small-scale enhancement of the clustering amplitude of DESI ELGs is aligned with the results of simulations focusing on galaxy mergers \citep[e.g.,][]{Wetzel2009}.

In addition to the behavior of spatial clustering properties of DESI ELGs, \citet{Rocher2023} find that DESI ELGs have an unexpected clustering property in velocity space. They find that on average, the velocity dispersion of the satellite DESI ELGs is larger than the velocity dispersion of dark matter particles by $\sim 30\%$. We argue that this observed property is likely associated with spectral profiles of DESI ELGs having two velocity components along the sightlines. As shown in Figure~\ref{fig:all_distribution}, for two redshift systems, the median value of the line of sight velocity offsets is $\sim \rm 150\, km/s$. However, the current $Redrock$ pipeline typically determines the redshift in the middle of the two velocity components. In this case, this can possibly introduce a  $\rm \sim 75 \, km/s$ velocity offset of the central galaxy redshifts systematically and adds additional velocity dispersion in the relative velocities between centrals and satellites. 

Here we consider a simple case to quantify possible signals.  Assuming that the dark matter halo mass of ELGs is $\approx 10^{12}\, M_{\odot}$ \citep[e.g.,][]{Rocher2023} with the corresponding velocity dispersion of dark matter particles being $\rm \sim 100 \, km/s$, adding a systematic $\rm 75\, km/s$ offset in the velocity measurements will increase the estimated velocity dispersion by $25\%$, which is similar to the results in \citet{Rocher2023}. This demonstrates that this redshift determination effect can be responsible for at least part of the apparent excess velocity dispersion of the satellite galaxies and it needs to be taken into account for the clustering measurements in velocity space.

\section{Conclusions}
By performing PCA on $\sim 230,000$ spectra of ELGs at $0.6<z<1.63$ from DESI Early Data Release, we decomposed [OII] profiles based on the derived PCA eigen-spectra and explored the diversity of [OII] line profiles in low-dimensional coefficient space. We further utilized the physical and morphological properties of galaxies in the COSMOS field and investigated the relationship between [OII] line profiles and the properties of the galaxies. 
Our main findings are summarized as follows:

\begin{itemize}
    \item We find that the first two eigen-spectra, which correspond to the line width (1st) and the peakedness (2nd) of [OII] doublet lines, can explain $\sim54\%$ of the total variance of [OII] line profiles. Using the coefficients of these two eigen-spectra, we show that DESI ELGs can be classified into at least three types with narrow [OII] lines, broad [OII] lines, and two-redshift systems, demonstrating the diversity of [OII] line profiles. 

    \item Combining PCA results with the galaxy physical properties from the COSMOS2020 catalog, we find that ELGs with broader line profiles tend to have higher stellar mass and SFR. Moreover, by fixing stellar mass, we find that ELGs with broader line profiles have higher median SFR. 
    We also find that DESI ELGs preferentially have higher SFR than the average SFR of the star-forming galaxies at the similar redshifts. These trends are observed across the entire redshift range of DESI ELGs from $z\sim0.6$ to $1.63$. 
    
    \item We include the morphological properties of DESI ELGs derived from HST ACS images and quantify the enhancement of various properties, including SFR and shape properties, of DESI ELGs with respect to the reference star-forming galaxies with similar stellar mass, sizes, and redshifts. We find that $\rm \Delta log_{10} SFR$ correlates the most with $\rm \Delta A$ and $\rm Coeff_{1}$ with Spearman's correlation coefficients with 0.31 and 0.16 respectively. Moreover, $\rm Coeff_{1}$ has the highest correlation coefficient with $\rm \Delta A$  ($\rho_{s}=0.08$ with $\sim 4 \sigma$ detection) than with other morphological parameters. 

    \item The median $\rm \Delta A$ and $\rm \Delta log_{10}\,SFR$ and the fraction of high $\rm \Delta A$ and high $\rm \Delta log_{10}\, SFR$ as a function of $\rm Coeff_{1}$ and $\rm Coeff_{2}$ both show that ELGs with broad line profiles have higher $\rm \Delta A$ and $\rm \Delta log_{10}\, SFR$ than ELGs with narrow or double-peak line profiles. This result reveals an underlying relationship between $\rm \Delta A$, $\rm \Delta log_{10}\,SFR$, and line profiles. 

    \item We show that ELGs with high $\rm \Delta A$ have on average $\sim0.2$ dex enhancement of the SFR than ELGs with low $\rm \Delta A$ and the $\rm Coeff_{1}$ can further contribute to the enhancement of SFR by $\sim0.1$ dex. 
\end{itemize}
Finally, we argue that this inter-relationship between the line profiles,  physical and morphological properties of DESI ELGs can be naturally explained by both the galaxy merger and disk instability star-forming clumps scenarios.

The results of this work show that the large DESI spectroscopic dataset opens a new window for statistically investigating the galaxy physical properties and how the mechanism drives galaxy evolution at $1<z<1.6$. 
The combination of the DESI data and upcoming imaging datasets provided by space telescopes, such as Euclid \citep{Euclid2022} and the Roman Space Telescope \citep{Akeson2019}, will further offer a large galaxy sample with both kinematics and morphological information of galaxies for galaxy evolution science. 

Understanding how the physical properties of galaxies link to the properties of dark matter halos and the large-scale environments is crucial for obtaining precise cosmological measurements. 
Recent galaxy clustering results of DESI ELGs indicate that the standard galaxy-halo connection models are insufficient to describe small-scale clustering measurements in both spatial and velocity space. As demonstrated in this work, this  is possibly due to the nature of DESI ELGs and the redshift determination effect due to their diverse line profiles.
It will be informative to perform clustering measurements as a function of emission line profiles and obtain a better understanding of small-scale clustering properties of the DESI ELG population. This approach can be used to utilize the hidden spectral information from the ongoing and upcoming cosmological spectroscopic surveys, including PFS \citep{Takada2014}, Euclid \citep{Euclid2022}, the Roman Space Telescope \citep{Akeson2019}, and the  next-generation surveys, e.g., DESI-II and Spec-5 experiments \citep{Schlegel2022}, that use emission line galaxies as primary tracers for the large-scale structure of the Universe.

\section*{Data Availability}
All data points shown in the figures are available at \href{https://zenodo.org/doi/10.5281/zenodo.13148442}{Zenodo}.

\begin{acknowledgments}
We thank the anonymous referee for the constructive report. 
We also want to thank Anand Raichoor and Chiara Circosta for their comments and suggestions for the early version of this paper and John Weaver for his help with using the COSMOS2020 catalog. TWL thanks Chian-Chou Chen for discussions and suggestions. TWL was supported by the National Science and Technology Council (MOST 111-2112-M-002-015-MY3, NSTC 113-2112-M-002-028-MY3), the Ministry of Education, Taiwan (MOE Yushan Young Scholar grant NTU-110VV007, NTU-111V1007-2, NTU-112V1007-3, NTU-113V1007-4), National Taiwan University research grant (NTU-CC-111L894806, NTU-CC-112L893606, NTU-CC-113L891806, NTU-111L7318, NTU-112L7302). JM gratefully acknowledges funding support for this work from the U.S. Department of Energy, Office of Science, Office of High Energy Physics under Award Number DE-SC0020086. M.S. acknowledges support by the Polish National Agency for Academic Exchange (Bekker grant BPN/BEK/2021/1/00298/DEC/1), the State Research Agency of the Spanish Ministry of Science and Innovation under the grants 'Galaxy Evolution with Artificial Intelligence' (PGC2018-100852-A-I00) and 'BASALT' (PID2021-126838NB-I00). This work was partially supported by the European Union's Horizon 2020 Research and Innovation program under the Maria Sklodowska-Curie grant agreement (No. 754510).

This material is based upon work supported by the U.S. Department of Energy (DOE), Office of Science, Office of High-Energy Physics, under Contract No. DE–AC02–05CH11231, and by the National Energy Research Scientific Computing Center, a DOE Office of Science User Facility under the same contract. Additional support for DESI was provided by the U.S. National Science Foundation (NSF), Division of Astronomical Sciences under Contract No. AST-0950945 to the NSF’s National Optical-Infrared Astronomy Research Laboratory; the Science and Technology Facilities Council of the United Kingdom; the Gordon and Betty Moore Foundation; the Heising-Simons Foundation; the French Alternative Energies and Atomic Energy Commission (CEA); the National Council of Humanities, Science and Technology of Mexico (CONAHCYT); the Ministry of Science and Innovation of Spain (MICINN), and by the DESI Member Institutions: \url{https://www.desi.lbl.gov/collaborating-institutions}.

The DESI Legacy Imaging Surveys consist of three individual and complementary projects: the Dark Energy Camera Legacy Survey (DECaLS), the Beijing-Arizona Sky Survey (BASS), and the Mayall z-band Legacy Survey (MzLS). DECaLS, BASS and MzLS together include data obtained, respectively, at the Blanco telescope, Cerro Tololo Inter-American Observatory, NSF’s NOIRLab; the Bok telescope, Steward Observatory, University of Arizona; and the Mayall telescope, Kitt Peak National Observatory, NOIRLab. NOIRLab is operated by the Association of Universities for Research in Astronomy (AURA) under a cooperative agreement with the National Science Foundation. Pipeline processing and analyses of the data were supported by NOIRLab and the Lawrence Berkeley National Laboratory. Legacy Surveys also uses data products from the Near-Earth Object Wide-field Infrared Survey Explorer (NEOWISE), a project of the Jet Propulsion Laboratory/California Institute of Technology, funded by the National Aeronautics and Space Administration. Legacy Surveys was supported by: the Director, Office of Science, Office of High Energy Physics of the U.S. Department of Energy; the National Energy Research Scientific Computing Center, a DOE Office of Science User Facility; the U.S. National Science Foundation, Division of Astronomical Sciences; the National Astronomical Observatories of China, the Chinese Academy of Sciences and the Chinese National Natural Science Foundation. LBNL is managed by the Regents of the University of California under contract to the U.S. Department of Energy. The complete acknowledgments can be found at \url{https://www.legacysurvey.org/}.

Any opinions, findings, and conclusions or recommendations expressed in this material are those of the author(s) and do not necessarily reflect the views of the U. S. National Science Foundation, the U. S. Department of Energy, or any of the listed funding agencies.

The authors are honored to be permitted to conduct scientific research on Iolkam Du’ag (Kitt Peak), a mountain with particular significance to the Tohono O’odham Nation.
\end{acknowledgments}

\software{Astropy \citep{astropy2013,astropy2018,astropy2022}, Numpy \citep{numpy}, Scipy \citep{scipy}, Matplotlib \citep{Hunter2007}, AstroML \citep{astroml}, KapteynPackage \citep{kmpfit}}

\newpage

\appendix
\section{Removing active galactic nuclei}
\label{appendix:AGN}
It is possible that a fraction of DESI ELGs having actively accreting 
supermassive black holes (active galactic nuclei, AGN) which can produce [OII] lines. To eliminate the complexity due to the AGN contributions, we utilize available line information obtained via the \textit{FastSpecFit}\footnote{\url{https://fastspecfit.readthedocs.io/en/latest/index.html}} algorithm developed by \citet{Moustakas2024}. We identify DESI ELGs that satisfy \textit{any} of the following conditions and remove them from our analysis:
\begin{equation}
    \rm log_{10}\, F_{[OIII]\lambda5007}/F_{H_{\beta}}>0.2\ \ \& \ \ line \ width_{[OIII]\lambda5007} \ FWHM>300 \, km/s,
\label{eq:type2}
\end{equation}
\begin{equation}
    \rm \frac{S}{N} (F_{MgII\lambda2796})>2, 
\label{eq:type1}
\end{equation}
\begin{equation}
    \rm \frac{S}{N} (F_{[NeV]\lambda3426})>2.
\label{eq:type22}
\end{equation}

The first condition \ref{eq:type2} follows the condition used in \citet{Zakamska2003} and \citet{Reyes2008} for selecting Type II quasars at $0.4<z<0.9$ where only $\rm [OIII]\lambda5007$ and $\rm H_{\beta}$ lines are accessible in the optical wavelength coverage. In \citet{Zakamska2003}, the selections are $\rm log_{10}\, F_{[OIII]\lambda5007}/F_{H_{\beta}}>0.3$ and  $\rm line \ width_{[OIII]5007} \ FWHM>400 \, km/s$. In our condition, we reduce the line ratio by 0.1 dex and the line width by 100 km/s so that our selection is conservative. 

The second condition, the signal-to-noise cut of MgII emission line  (\ref{eq:type1}), is used to remove ELG spectra with possible broad MgII emission lines originated from quasars, e.g., the missing quasar population in \citet{Alexander2023}. 

The third condition, the signal-to-noise cut of $[NeV]\lambda3426$ emission line (\ref{eq:type22}), is motivated by the fact that the existence of $\rm [NeV]\lambda3426$ line requires relatively energetic radiation and AGN has been considered as the main source for producing the required radiation \citep[e.g.,][]{Maddox2018, Cleri2023}. 

Approximately $4.3\%$ of the total sources in the original DESI ELG sample meet the above conditions. Individually, \ref{eq:type2}, \ref{eq:type1}, and \ref{eq:type22} contribute to $\sim0.7\%$, $\sim 0.9\%$, and $\sim 3.0\%$ respectively. The final $4.3\%$ includes sources in common.

\section{Morphology parameters as a function of line profiles} 
\label{appendix:morphology}
\begin{figure*}
\center
\includegraphics[width=1\textwidth]{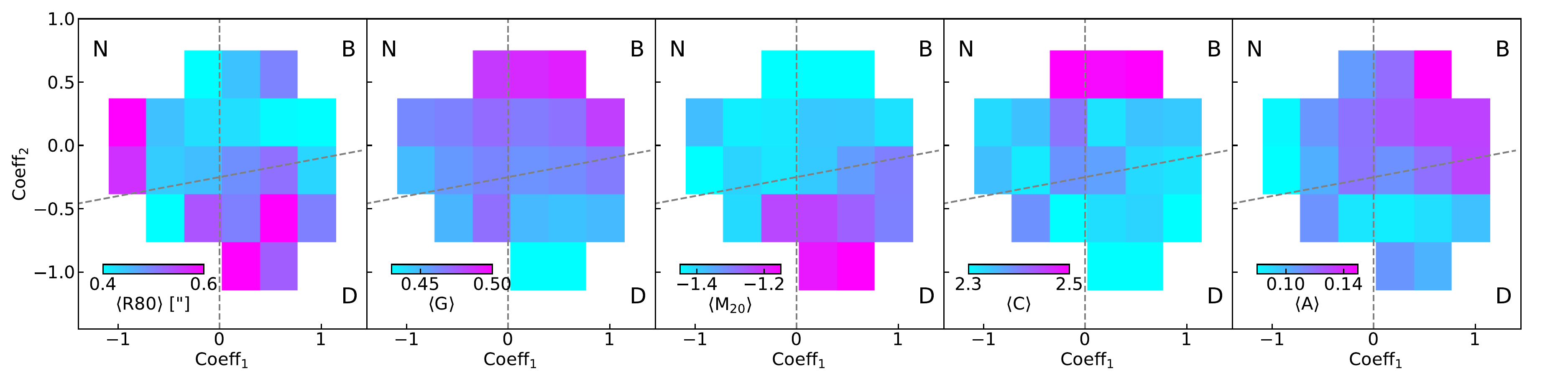}
\caption{Morphological parameters as a function of $\rm Coeff_{1}$ and $\rm Coeff_{2}$. Median values of galaxy size (R80 in unit of arcsec) , Gini coefficient($\rm G$), second-order moment of the brightest $20\%$ pixels (M20), concentration (C), and asymmetry (A) are shown from left to right respectively.}
\label{fig:morphology}
\end{figure*}
Here we summarize the relationships between the morphological parameters and the line profiles. Figure~\ref{fig:morphology} shows the median values of the morphological parameters, galaxy size (R80),  Gini coefficient (G), the second-order moment of the brightest $20\%$ pixels (M20), concentration (C), and asymmetry (A) from left to right respectively, as a function of $\rm Coeff_{1}$ and $\rm Coeff_{2}$. These results are based on the DESI-ZEST sample.

Trends can be observed from Figure~\ref{fig:morphology}. First, DESI ELGs in the double-peak region tend to have larger sizes than DESI ELGs in other regions as shown in the first panel. This is consistent with the fact that DESI ELGs in the double-peak region also have higher stellar mass. Correlating with R80, the M20 values also tend to be larger in the double-peak region. We note that the DESI ELGs with $\rm Coeff_{1}\sim-1$ have on average lower redshifts and therefore have larger observed sizes. 
Second, Gini coefficient (G) and asymmetry parameter (A) behave similarly with higher median values in the broad region, indicating that DESI ELGs in the broad region tend to have relatively disturbed morphology. Finally, the concentration C anti-correlates with M20, having the highest median values for DESI ELGs with high $\rm Coeff_{2}$. These internal correlations between parameters are consistent with the results summarized in \citet{COSMOS_zbest}. We note that while these morphological parameters all have some dependencies with the line profiles, we focus on the asymmetry parameter in this work given that the excess asymmetry parameter $\rm \Delta A$ yields the highest Spearman's correlation coefficients with $\rm \Delta log_{10}\, SFR$ and $\rm Coeff_{1}$ and $\rm Coeff_{2}$ as shown in Section 4.


\begin{thebibliography}{}
\bibitem[Abraham et al.(2003)]{Abraham2003} Abraham, R.~G., van den Bergh, S., \& Nair, P.\ 2003, \apj, 588, 218. doi:10.1086/373919

\bibitem[Aihara et al.(2018)]{Aihara2018} Aihara, H., Arimoto, N., Armstrong, R., et al.\ 2018, \pasj, 70, S4. doi:10.1093/pasj/psx066

\bibitem[Aihara et al.(2019)]{Aihara2019} Aihara, H., AlSayyad, Y., Ando, M., et al.\ 2019, \pasj, 71, 114. doi:10.1093/pasj/psz103

\bibitem[Alexander et al.(2023)]{Alexander2023} Alexander, D.~M., Davis, T.~M., Chaussidon, E., et al.\ 2023, \aj, 165, 124. doi:10.3847/1538-3881/acacfc
\bibitem[Anand et al.(2024)]{Anand2024} Anand, A., Guy, J., Bailey, S., et al.\ 2024, arXiv:2405.19288. doi:10.48550/arXiv.2405.19288

\bibitem[Akeson et al.(2019)]{Akeson2019} Akeson, R., Armus, L., Bachelet, E., et al.\ 2019, arXiv:1902.05569. doi:10.48550/arXiv.1902.05569

\bibitem[Astropy Collaboration et al.(2013)]{astropy2013} Astropy Collaboration, Robitaille, T.~P., Tollerud, E.~J., et al.\ 2013, \aap, 558, A33. doi:10.1051/0004-6361/201322068

\bibitem[Astropy Collaboration et al.(2018)]{astropy2018} Astropy Collaboration, Price-Whelan, A.~M., Sip{\H{o}}cz, B.~M., et al.\ 2018, \aj, 156, 123. doi:10.3847/1538-3881/aabc4f

\bibitem[Astropy Collaboration et al.(2022)]{astropy2022} Astropy Collaboration, Price-Whelan, A.~M., Lim, P.~L., et al.\ 2022, \apj, 935, 167. doi:10.3847/1538-4357/ac7c74


\bibitem[Bailey(2012)]{Bailey2012} Bailey, S.\ 2012, \pasp, 124, 1015. doi:10.1086/668105

\bibitem[Bailey et al.(2023)]{Bailey2023} Bailey, S. et al. in prep

\bibitem[Behroozi et al.(2015)]{Behroozi2015} Behroozi, P.~S., Zhu, G., Ferguson, H.~C., et al.\ 2015, \mnras, 450, 1546. doi:10.1093/mnras/stv728

\bibitem[Bertin \& Arnouts(1996)]{Bertin1996} Bertin, E. \& Arnouts, S.\ 1996, \aaps, 117, 393. doi:10.1051/aas:1996164

\bibitem[Blake et al.(2011)]{Blake2011} Blake, C., Kazin, E.~A., Beutler, F., et al.\ 2011, \mnras, 418, 1707. doi:10.1111/j.1365-2966.2011.19592.x

\bibitem[Bottrell et al.(2023)]{Bottrell2023} Bottrell, C., Yesuf, H.~M., Popping, G., et al.\ 2023, \mnras. doi:10.1093/mnras/stad2971


\bibitem[Brammer et al.(2008)]{Brammer2008} Brammer, G.~B., van Dokkum, P.~G., \& Coppi, P.\ 2008, \apj, 686, 1503. doi:10.1086/591786
\bibitem[Brodzeller et al.(2023)]{Brodzeller2023} Brodzeller, A., Dawson, K., Bailey, S., et al.\ 2023, \aj, 166, 66. doi:10.3847/1538-3881/ace35d

\bibitem[Chabrier(2003)]{Chabrier2003} Chabrier, G.\ 2003, \pasp, 115, 763. doi:10.1086/376392

\bibitem[Chaussidon et al.(2023)]{DESI_QSO} Chaussidon, E., Y{\`e}che, C., Palanque-Delabrouille, N., et al.\ 2023, \apj, 944, 107. doi:10.3847/1538-4357/acb3c2
\bibitem[Chen et al.(2016)]{Chen2016} Chen, Y.-M., Gu, Q.-S., Tremonti, C.~A., et al.\ 2016, \mnras, 459, 3861. doi:10.1093/mnras/stw942
\bibitem[Cleri et al.(2023)]{Cleri2023} Cleri, N.~J., Yang, G., Papovich, C., et al.\ 2023, \apj, 948, 112. doi:10.3847/1538-4357/acc1e6

\bibitem[Comparat et al.(2013)]{Comparat2013} Comparat, J., Kneib, J.-P., Bacon, R., et al.\ 2013, \aap, 559, A18. doi:10.1051/0004-6361/201322452

\bibitem[Comparat et al.(2015)]{Comparat2015} Comparat, J., Richard, J., Kneib, J.-P., et al.\ 2015, \aap, 575, A40. doi:10.1051/0004-6361/201424767

\bibitem[Conselice(2014)]{Conselice2014} Conselice, C.~J.\ 2014, \araa, 52, 291. doi:10.1146/annurev-astro-081913-040037

\bibitem[Cooper et al.(2023)]{DESI_MW} Cooper, A.~P., Koposov, S.~E., Allende Prieto, C., et al.\ 2023, \apj, 947, 37. doi:10.3847/1538-4357/acb3c0

\bibitem[COSMOS team (2007)]{cosmos_ACS} COSMOS team 2007, COSMOS Zurich Structure \& Morphology Catalog, v1, IPAC, doi:10.26131/IRSA160

\bibitem[Dawson et al.(2016)]{DawsoneBOSS} Dawson, K.~S., Kneib, J.-P., Percival, W.~J., et al.\ 2016, \aj, 151, 44. doi:10.3847/0004-6256/151/2/44

\bibitem[Delchambre(2015)]{Delchambre2015} Delchambre, L.\ 2015, \mnras, 446, 3545. doi:10.1093/mnras/stu2219

\bibitem[Dekel et al.(2009)]{Dekel2009} Dekel, A., Sari, R., \& Ceverino, D.\ 2009, \apj, 703, 785. doi:10.1088/0004-637X/703/1/785

\bibitem[DESI Collaboration et al.(2016a)]{DESI2016} DESI Collaboration, Aghamousa, A., Aguilar, J., et al.\ 2016, arXiv:1611.00036. doi:10.48550/arXiv.1611.00036

\bibitem[DESI Collaboration et al.(2016b)]{DESI2016b} DESI Collaboration, Aghamousa, A., Aguilar, J., et al.\ 2016, arXiv:1611.00037. doi:10.48550/arXiv.1611.00037

\bibitem[DESI Collaboration et al.(2022)]{DESI_overview} DESI Collaboration, Abareshi, B., Aguilar, J., et al.\ 2022, \aj, 164, 207. doi:10.3847/1538-3881/ac882b

\bibitem[DESI Collaboration et al.(2024a)]{DESI_SV} DESI Collaboration, Adame, A.~G., Aguilar, J., et al.\ 2024, \aj, 167, 62. doi:10.3847/1538-3881/ad0b08

\bibitem[DESI Collaboration et al.(2024b)]{DESI_EDR} DESI Collaboration, Adame, A.~G., Aguilar, J., et al.\ 2024, \aj, 168, 58. doi:10.3847/1538-3881/ad3217

\bibitem[Dey et al.(2019)]{Dey2019} Dey, A., Schlegel, D.~J., Lang, D., et al.\ 2019, \aj, 157, 168. doi:10.3847/1538-3881/ab089d


\bibitem[Drinkwater et al.(2010)]{Drinkwater2010} Drinkwater, M.~J., Jurek, R.~J., Blake, C., et al.\ 2010, \mnras, 401, 1429. doi:10.1111/j.1365-2966.2009.15754.x

\bibitem[Eisenstein et al.(2005)]{Eisenstein2005} Eisenstein, D.~J., Zehavi, I., Hogg, D.~W., et al.\ 2005, \apj, 633, 560. doi:10.1086/466512
\bibitem[Euclid Collaboration et al.(2022)]{Euclid2022} Euclid Collaboration, Scaramella, R., Amiaux, J., et al.\ 2022, \aap, 662, A112. doi:10.1051/0004-6361/202141938

\bibitem[Ellison et al.(2008)]{Ellison2008} Ellison, S.~L., Patton, D.~R., Simard, L., et al.\ 2008, \aj, 135, 1877. doi:10.1088/0004-6256/135/5/1877
\bibitem[Ellison et al.(2010)]{Ellison2010} Ellison, S.~L., Patton, D.~R., Simard, L., et al.\ 2010, \mnras, 407, 1514. doi:10.1111/j.1365-2966.2010.17076.x

\bibitem[Elmegreen et al.(2021)]{Elmegreen2021} Elmegreen, D.~M., Elmegreen, B.~G., Whitmore, B.~C., et al.\ 2021, \apj, 908, 121. doi:10.3847/1538-4357/abd541


\bibitem[Fisher et al.(2017)]{Fisher2017} Fisher, D.~B., Glazebrook, K., Abraham, R.~G., et al.\ 2017, \apjl, 839, L5. doi:10.3847/2041-8213/aa6478

\bibitem[Gao et al.(2023)]{Gao2023} Gao, H., Jing, Y.~P., Gui, S., et al.\ 2023, \apj, 954, 207. doi:10.3847/1538-4357/ace90a

\bibitem[Gao et al.(2024)]{Gao2024} Gao, H., Jing, Y.~P., Xu, K., et al.\ 2024, \apj, 961, 74. doi:10.3847/1538-4357/ad09d6

\bibitem[Goldbaum et al.(2016)]{Goldbaum2016} Goldbaum, N.~J., Krumholz, M.~R., \& Forbes, J.~C.\ 2016, \apj, 827, 28. doi:10.3847/0004-637X/827/1/28

\bibitem[Gould et al.(2023)]{cosmos_eazy} Gould, K.~M.~L., Brammer, G., Valentino, F., et al.\ 2023, \aj, 165, 248. doi:10.3847/1538-3881/accadc
\bibitem[Guo et al.(2015)]{Guo2015} Guo, Y., Ferguson, H.~C., Bell, E.~F., et al.\ 2015, \apj, 800, 39. doi:10.1088/0004-637X/800/1/39

\bibitem[Guy et al.(2023)]{Guy2023} Guy, J., Bailey, S., Kremin, A., et al.\ 2023, \aj, 165, 144. doi:10.3847/1538-3881/acb212

\bibitem[Hahn et al.(2023)]{DESI_BGS} Hahn, C., Wilson, M.~J., Ruiz-Macias, O., et al.\ 2023, \aj, 165, 253. doi:10.3847/1538-3881/accff8

\bibitem[Harris et al. (2020)]{numpy} Harris, C.R., Millman, K.J., van der Walt, S.J. et al. \ 2020, Nature 585, 357–362. doi: 10.1038/s41586-020-2649-2

\bibitem[Hinshaw et al.(2013)]{WMAP9} Hinshaw, G., Larson, D., Komatsu, E., et al.\ 2013, \apjs, 208, 19. doi:10.1088/0067-0049/208/2/19

\bibitem[Hopkins et al.(2008)]{Hopkins2008} Hopkins, P.~F., Hernquist, L., Cox, T.~J., et al.\ 2008, \apjs, 175, 356. doi:10.1086/524362

\bibitem[Hunter (2007)]{Hunter2007} Hunter, J. D. 2007, Computing in Science \& Engineering, 9, 90. doi:10.1109/MCSE.2007.55

\bibitem[Ilbert et al.(2006)]{Ilbert2006} Ilbert, O., Arnouts, S., McCracken, H.~J., et al.\ 2006, \aap, 457, 841. doi:10.1051/0004-6361:20065138


\bibitem[Jolliffe (2002)]{PCA_review} Jolliffe, Ian,
2002, Principal component analysis, 338-372, Publisher Springer New York
\bibitem[Kassin et al.(2012)]{Kassin2012} Kassin, S.~A., Weiner, B.~J., Faber, S.~M., et al.\ 2012, \apj, 758, 106. doi:10.1088/0004-637X/758/2/106
\bibitem[Kaasinen et al.(2017)]{COSMOS_OII} Kaasinen, M., Bian, F., Groves, B., et al.\ 2017, \mnras, 465, 3220. doi:10.1093/mnras/stw2827

\bibitem[Kennicutt(1998)]{Kennicutt1998} Kennicutt, R.~C.\ 1998, \araa, 36, 189. doi:10.1146/annurev.astro.36.1.189
\bibitem[Kennicutt et al.(2009)]{Kennicutt2009} Kennicutt, R.~C., Hao, C.-N., Calzetti, D., et al.\ 2009, \apj, 703, 1672. doi:10.1088/0004-637X/703/2/1672


\bibitem[Kewley et al.(2019)]{Kewley2019} Kewley, L.~J., Nicholls, D.~C., \& Sutherland, R.~S.\ 2019, \araa, 57, 511. doi:10.1146/annurev-astro-081817-051832


\bibitem[Koekemoer et al.(2007)]{Koekemoer2007} Koekemoer, A.~M., Aussel, H., Calzetti, D., et al.\ 2007, \apjs, 172, 196. doi:10.1086/520086
\bibitem[Kriek \& Conroy(2013)]{Kriek2013} Kriek, M. \& Conroy, C.\ 2013, \apjl, 775, L16. doi:10.1088/2041-8205/775/1/L16

\bibitem[Lan et al.(2023)]{Lan2023} Lan, T.-W., Tojeiro, R., Armengaud, E., et al.\ 2023, \apj, 943, 68. doi:10.3847/1538-4357/aca5fa

\bibitem[Lang et al.(2016)]{Lang2016} Lang, D., Hogg, D.~W., \& Mykytyn, D.\ 2016, Astrophysics Source Code Library. ascl:1604.008

\bibitem[Law et al.(2022)]{Law2022} Law, D.~R., Belfiore, F., Bershady, M.~A., et al.\ 2022, \apj, 928, 58. doi:10.3847/1538-4357/ac5620

\bibitem[Leauthaud et al.(2007)]{Leauthaud2007} Leauthaud, A., Massey, R., Kneib, J.-P., et al.\ 2007, \apjs, 172, 219. doi:10.1086/516598
\bibitem[Levi et al.(2013)]{Levi2013} Levi, M., Bebek, C., Beers, T., et al.\ 2013, arXiv:1308.0847

\bibitem[Lotz et al.(2004)]{Lotz2004} Lotz, J.~M., Primack, J., \& Madau, P.\ 2004, \aj, 128, 163. doi:10.1086/421849

\bibitem[Lotz et al.(2008)]{Lotz2008} Lotz, J.~M., Jonsson, P., Cox, T.~J., et al.\ 2008, \mnras, 391, 1137. doi:10.1111/j.1365-2966.2008.14004.x
\bibitem[Madau \& Dickinson(2014)]{Madau2014} Madau, P. \& Dickinson, M.\ 2014, \araa, 52, 415. doi:10.1146/annurev-astro-081811-125615
\bibitem[Maddox (2018)]{Maddox2018} Maddox, N.\ 2018, \mnras, 480, 5203. doi:10.1093/mnras/sty2201

\bibitem[Mai et al.(2024)]{Mai2024} Mai, Y., Croom, S.~M., Wisnioski, E., et al.\ 2024, \mnras. doi:10.1093/mnras/stae2033

\bibitem[Mandelker et al.(2014)]{Mandelker2014} Mandelker, N., Dekel, A., Ceverino, D., et al.\ 2014, \mnras, 443, 3675. doi:10.1093/mnras/stu1340
\bibitem[Martin et al.(2023)]{Martin2023} Martin, A., Guo, Y., Wang, X., et al.\ 2023, \apj, 955, 106. doi:10.3847/1538-4357/aced3e

\bibitem[Maschmann et al.(2020)]{Maschmann2020} Maschmann, D., Melchior, A.-L., Mamon, G.~A., et al.\ 2020, \aap, 641, A171. doi:10.1051/0004-6361/202037868

\bibitem[Maschmann et al.(2023)]{Maschmann2023} Maschmann, D., Halle, A., Melchior, A.-L., et al.\ 2023, \aap, 670, A46. doi:10.1051/0004-6361/202244746




\bibitem[McCracken et al.(2012)]{McCracken2012} McCracken, H.~J., Milvang-Jensen, B., Dunlop, J., et al.\ 2012, \aap, 544, A156. doi:10.1051/0004-6361/201219507

\bibitem[Miller et al.(2023)]{Miller2023} Miller, T.~N., Doel, P., Gutierrez, G., et al.\ 2023, arXiv:2306.06310. doi:10.48550/arXiv.2306.06310

\bibitem[Moustakas et al.(2006)]{Moustakas2006} Moustakas, J., Kennicutt, R.~C., \& Tremonti, C.~A.\ 2006, \apj, 642, 775. doi:10.1086/500964

\bibitem[Moustakas et al.(2013)]{Moustakas2013} Moustakas, J., Coil, A.~L., Aird, J., et al.\ 2013, \apj, 767, 50. doi:10.1088/0004-637X/767/1/50
\bibitem[Moustakas et al.(2023)]{Moustakas2023} Moustakas, J., Lang, D., Dey, A., et al.\ 2023, \apjs, 269, 3. doi:10.3847/1538-4365/acfaa2

\bibitem[Moustakas et al.(2024)]{Moustakas2024} Moustakas, J. et al. in prep. 



\bibitem[Murata et al.(2014)]{Murata2014} Murata, K.~L., Kajisawa, M., Taniguchi, Y., et al.\ 2014, \apj, 786, 15. doi:10.1088/0004-637X/786/1/15

\bibitem[Myers et al.(2023)]{Myers2023} Myers, A.~D., Moustakas, J., Bailey, S., et al.\ 2023, \aj, 165, 50. doi:10.3847/1538-3881/aca5f9

\bibitem[Oliva-Altamirano et al.(2018)]{Altamirano2018} Oliva-Altamirano, P., Fisher, D.~B., Glazebrook, K., et al.\ 2018, \mnras, 474, 522. doi:10.1093/mnras/stx2797

\bibitem[Osterbrock \& Ferland(2006)]{OsterbrockAGN} Osterbrock, D.~E. \& Ferland, G.~J.\ 2006, Astrophysics of gaseous nebulae and active galactic nuclei, 2nd. ed. by D.E. Osterbrock and G.J. Ferland. Sausalito, CA: University Science Books, 2006

\bibitem[Patton et al.(2020)]{Patton2020} Patton, D.~R., Wilson, K.~D., Metrow, C.~J., et al.\ 2020, \mnras, 494, 4969. doi:10.1093/mnras/staa913

\bibitem[Raichoor et al.(2021)]{Raichoor2021} Raichoor, A., de Mattia, A., Ross, A.~J., et al.\ 2021, \mnras, 500, 3254. doi:10.1093/mnras/staa3336

\bibitem[Raichoor et al.(2023)]{DESI_ELG} Raichoor, A., Moustakas, J., Newman, J.~A., et al.\ 2023, \aj, 165, 126. doi:10.3847/1538-3881/acb213
\bibitem[Raichoor et al.(2024)]{Raichoor2024} Raichoor, A., et al. in prep.
\bibitem[Reyes et al.(2008)]{Reyes2008} Reyes, R., Zakamska, N.~L., Strauss, M.~A., et al.\ 2008, \aj, 136, 2373. doi:10.1088/0004-6256/136/6/2373

\bibitem[Ribeiro et al.(2017)]{Ribeiro2017} Ribeiro, B., Le F{\`e}vre, O., Cassata, P., et al.\ 2017, \aap, 608, A16. doi:10.1051/0004-6361/201630057

\bibitem[Rocher et al.(2023)]{Rocher2023} Rocher, A., Ruhlmann-Kleider, V., Burtin, E., et al.\ 2023, \jcap, 2023, 016. doi:10.1088/1475-7516/2023/10/016



\bibitem[Sattari et al.(2023)]{Sattari2023} Sattari, Z., Mobasher, B., Chartab, N., et al.\ 2023, \apj, 951, 147. doi:10.3847/1538-4357/acd5d6

\bibitem[Scarlata et al.(2007)]{COSMOS_zbest} Scarlata, C., Carollo, C.~M., Lilly, S., et al.\ 2007, \apjs, 172, 406. doi:10.1086/516582

\bibitem[Schlafly et al.(2023)]{Schlafly2023} Schlafly, E.~F., Kirkby, D., Schlegel, D.~J., et al.\ 2023, arXiv:2306.06309. doi:10.48550/arXiv.2306.06309

\bibitem[Schlegel et al.(1998)]{Schlegel1998} Schlegel, D.~J., Finkbeiner, D.~P., \& Davis, M.\ 1998, \apj, 500, 525. doi:10.1086/305772


\bibitem[Schlegel et al.(2022)]{Schlegel2022} Schlegel, D.~J., Ferraro, S., Aldering, G., et al.\ 2022, arXiv:2209.03585. doi:10.48550/arXiv.2209.03585


\bibitem[Scoville et al.(2007)]{Scoville2007} Scoville, N., Abraham, R.~G., Aussel, H., et al.\ 2007, \apjs, 172, 38. doi:10.1086/516580

\bibitem[Scoville et al.(2007)]{Scoville2007b} Scoville, N., Aussel, H., Brusa, M., et al.\ 2007, \apjs, 172, 1. doi:10.1086/516585

\bibitem[Silber et al.(2023)]{Silber2023} Silber, J.~H., Fagrelius, P., Fanning, K., et al.\ 2023, \aj, 165, 9. doi:10.3847/1538-3881/ac9ab1

\bibitem[Suzuki(2006)]{SuzukiPCA} Suzuki, N.\ 2006, \apjs, 163, 110. doi:10.1086/499272

\bibitem[Taghizadeh-Popp et al.(2012)]{PCAgalaxu} Taghizadeh-Popp, M., Heinis, S., \& Szalay, A.~S.\ 2012, \apj, 755, 143. doi:10.1088/0004-637X/755/2/143
\bibitem[Takada et al.(2014)]{Takada2014} Takada, M., Ellis, R.~S., Chiba, M., et al.\ 2014, \pasj, 66, R1. doi:10.1093/pasj/pst019
\bibitem[Terlouw and Vogelaar (2014)]{kmpfit} Terlouw, J.P. and Vogelaar, M.~G.~R. \ 2014, \url{http://www.astro.rug.nl/software/kapteyn/}

\bibitem[{\"U}bler et al.(2019)]{Ubler2019} {\"U}bler, H., Genzel, R., Wisnioski, E., et al.\ 2019, \apj, 880, 48. doi:10.3847/1538-4357/ab27cc

\bibitem[Vanderplas et al. (2012)]{astroml} Vanderplas, J.T., Connolly, A.J., Ivezi{\'c}, {\v Z}. and Gray, A., proc. of CIDU, pp. 47-54, 2012.  doi:10.1109/CIDU.2012.6382200

\bibitem[van der Wel et al.(2014)]{Wel2014} van der Wel, A., Franx, M., van Dokkum, P.~G., et al.\ 2014, \apj, 788, 28. doi:10.1088/0004-637X/788/1/28

\bibitem[Virtanen et al. (2020)]{scipy} Virtanen, P., Gommers, R., Oliphant, T.E., et al. \ 2020, Nature Methods, 17(3), 261-272. doi: 10.1038/s41592-019-0686-2

\bibitem[Yesuf et al.(2021)]{Yesuf2021} Yesuf, H.~M., Ho, L.~C., \& Faber, S.~M.\ 2021, \apj, 923, 205. doi:10.3847/1538-4357/ac27a7
\bibitem[Yip et al.(2004)]{YipPCA} Yip, C.~W., Connolly, A.~J., Szalay, A.~S., et al.\ 2004, \aj, 128, 585. doi:10.1086/422429

\bibitem[Yuan et al.(2023)]{Yuan2023} Yuan, S., Wechsler, R.~H., Wang, Y., et al.\ 2023, arXiv:2310.09329. doi:10.48550/arXiv.2310.09329


\bibitem[Weaver et al.(2022)]{COSMOS2020} Weaver, J.~R., Kauffmann, O.~B., Ilbert, O., et al.\ 2022, \apjs, 258, 11. doi:10.3847/1538-4365/ac3078

\bibitem[Wechsler \& Tinker(2018)]{Wechsler2018} Wechsler, R.~H. \& Tinker, J.~L.\ 2018, \araa, 56, 435. doi:10.1146/annurev-astro-081817-051756

\bibitem[Wetzel et al.(2009)]{Wetzel2009} Wetzel, A.~R., Cohn, J.~D., \& White, M.\ 2009, \mnras, 394, 2182. doi:10.1111/j.1365-2966.2009.14488.x
\bibitem[Wright et al.(2010)]{Wright2010} Wright, E.~L., Eisenhardt, P.~R.~M., Mainzer, A.~K., et al.\ 2010, \aj, 140, 1868. doi:10.1088/0004-6256/140/6/1868
\bibitem[Zakamska et al.(2003)]{Zakamska2003} Zakamska, N.~L., Strauss, M.~A., Krolik, J.~H., et al.\ 2003, \aj, 126, 2125. doi:10.1086/378610
\bibitem[Zhou et al.(2023)]{DESI_LRG} Zhou, R., Dey, B., Newman, J.~A., et al.\ 2023, \aj, 165, 58. doi:10.3847/1538-3881/aca5fb


\end{thebibliography}
\end{document}